\documentclass[aps,pra,freqn,reprint,showpacs,groupedaddress]{revtex4-1}
\usepackage{graphicx}

\begin{document}

\title{Proposed method for laser spectroscopy of pionic helium atoms to determine the charged-pion mass}
\author{Masaki Hori}
\affiliation{Max-Planck-Institut f\"{u}r Quantenoptik, Hans-Kopfermann-Strasse 1, 85748 Garching, Germany}
\author{Anna S\'{o}t\'{e}r}
\affiliation{Max-Planck-Institut f\"{u}r Quantenoptik, Hans-Kopfermann-Strasse 1, 85748 Garching, Germany}
\author{Vladimir I.~Korobov}
\affiliation{Joint Institute for Nuclear Research, 141980, Dubna, Russia}
\date{\today}
\pacs{36.10.Gv, 42.62.Fi, 14.40.Be}

\begin{abstract}
Metastable pionic helium ($\pi{\rm He}^+$) is a three-body atom composed of a helium nucleus, 
an electron occupying the $1s$ ground state, and a negatively charged pion $\pi^-$ in a Rydberg state with principal- 
and orbital angular momentum quantum numbers of $n\sim \ell+1\sim 16$. We calculate the spin-independent
energies of the $\pi{\rm ^3He}^+$ and $\pi{\rm ^4He}^+$ isotopes in the region $n=15$-19. These include
relativistic and quantum electrodynamics corrections of orders $R_{\infty}\alpha^2$ and $R_{\infty}\alpha^3$ in
atomic units, where $R_{\infty}$ and $\alpha$ denote the Rydberg and fine structure constants. The fine-structure 
splitting due to the coupling between the electron spin and the orbital angular momentum of the $\pi^-$, and the
radiative and Auger decay rates of the states are also calculated. Some states $(n,\ell)=(16,15)$ and $(17,16)$ retain 
nanosecond-scale lifetimes against $\pi^-$ absorption into the helium nucleus. We propose to use laser pulses
to induce $\pi^-$ transitions from these metastable states, to states with large ($\sim 10^{11}$ s$^{-1}$) 
Auger rates. The $\pi{\rm He}^{2+}$ ion that remains after Auger emission of the $1s$ electron undergoes Stark 
mixing with the $s$, $p$, and $d$ states during collisions with the helium atoms in the experimental target. 
This leads to immediate nuclear absorption of the $\pi^-$. The resonance condition between the laser beam and the atom is thus revealed as a sharp spike 
in the rates of neutrons, protons, deuterons, and tritons that emerge. A resonance curve is obtained 
from which the $\pi{\rm He}^+$ transition frequency can in principle be determined with a fractional precision of 
$10^{-8}-10^{-6}$ provided that the systematic uncertainties can be controlled. By comparing the 
measured $\pi{\rm He}^+$ frequencies with the calculated values, the $\pi^-$ mass may be determined with a 
similar precision. The $\pi{\rm He}^+$ will be synthesized by allowing a high-intensity ($>10^8$ s$^{-1}$)
beam of $\pi^-$　produced by a cyclotron to come to rest in a helium target. The precise time structure of the
$\pi^-$ beam is used to ensure a sufficient rate of coincidence between the resonant laser pulses and the 
$\pi{\rm He}^+$ atoms. 
\end{abstract}
\maketitle

\section{Introduction}
\label{intro} 

In this paper we describe a possible method for laser spectroscopy of metastable pionic helium 
($\pi{\rm He}^+\equiv\pi^{-}+{\rm He}^{2+}+e^-$). This is a hypothetical three-body atom 
\cite{condo1964,russell1969,russell1970_1,*russell1970_2,*russell1970_3} composed of a 
helium nucleus, an electron occupying the 1s ground state, and a negatively charged pion $\pi^-$ 
in a Rydberg state with principal and orbital angular momentum quantum numbers of around $n\sim \ell+1\sim 16$.
These states are theoretically expected to retain nanosecond-scale lifetimes against the competing processes of
$\pi^-$ absorption into the helium nucleus and 
$\pi^-\rightarrow\mu^- +\overline{\nu}_{\mu}$ decay to a negatively charged muon and a 
muon-based antineutrino. This is because the Rydberg $\pi^-$ orbitals have very little overlap with the nucleus, 
whereas the electromagnetic cascade processes that
normally cause the rapid deexcitation of the $\pi^-$, such as radiative decay, or 
Auger emission of the $1s$ electron, are relatively slow. 
The atom should therefore be amenable to laser spectroscopic measurements of the $\pi^-$ transition frequencies. 
This would conclusively show the existence of $\pi{\rm He}^+$. By comparing the experimental frequencies with 
the results of the three-body quantum electrodynamics (QED) calculations presented in this paper, the $\pi^-$ mass 
can in principle be determined with a high precision, as in the case of antiprotonic helium atoms 
\cite{mhori2011,korobov2008}. 

The existence of $\pi{\rm He}^+$ has been indirectly inferred from four experiments
\cite{fetkovich1963,block1963,block1965,zaimidoroga1967,nakamura1992} that observed that 
a small fraction of $\pi^-$ retains an anomalously long lifetime in helium targets.
Quantitative comparisons of the data with theoretical calculations have been difficult. Some sets of
calculated decay rates of $\pi{\rm He}^+$ states differ from each other by 1--2 orders of magnitude. 
Nothing is experimentally known about the 
distribution of states which may be formed. Whereas x-ray transitions between 
short-lived states of low principal quantum number $n_i$ in the two-body pionic helium
($\pi{\rm He}^{2+}\equiv \pi^- +{\rm He}^{2+}$) ion have been studied for many years
by fluorescence spectroscopy with an experimental precision of $\sim 10^{-4}$ 
\cite{wetmore1967,backenstoss1974,gotta2004}, no atomic lines of the three-body $\pi{\rm He}^+$ 
have been detected so far. Many assumptions are therefore needed to design any laser spectroscopy experiment.

We propose the irradiation of the $\pi{\rm He}^+$ with resonant laser pulses that induce transitions
from the metastable states to states with picosecond-scale lifetimes 
against Auger emission of the $1s$ electron. The Rydberg $\pi{\rm He}^{2+}$ ion that remains after Auger 
decay undergoes Stark mixing during collisions \cite{day1960,borie1980,landua1982} with the helium 
atoms in the experimental target. The electric fields induced by the collisions mix the ionic 
states with the $s$, $p$, and $d$ states at high $n_i$, that have large overlap
with the helium nucleus. This leads to nuclear absorption of the $\pi^-$ within 
picoseconds. Neutrons, protons, deuterons, and tritons with kinetic energies of up to 30--90 MeV 
consequently emerge. By measuring these particle rates as a function of the laser frequency, the 
resonance condition between the laser and the $\pi{\rm He}^+$ is revealed in the form of a resonance curve.
From this the $\pi{\rm He}^+$ transition frequency $\nu_{\rm exp}$ can be determined. 
The $\pi{\rm He}^+$ are synthesized by allowing
a $\pi^-$ beam produced by the Paul Scherrer Institute (PSI) ring cyclotron \cite{seidel2010,meg2013} to 
come to rest in a helium target. The precise time structure of this beam allows the 
formation of $\pi{\rm He}^+$ in the target to be synchronized to the arrival of the resonant 
laser pulses. Laser beams generated by solid-state lasers with high 
repetition rates $f_{\rm las}=0.1$--1 kHz and average powers 1--100 W will excite the $\pi{\rm He}^+$. 

The experiment is difficult for many reasons; in fact, laser excitation of a meson has
never been observed. The metastability of $\pi^-$ in helium corresponds to a lifetime 
$\tau_c\sim 7$ ns \cite{nakamura1992} which is shorter than that of any exotic atom studied by
laser spectroscopy so far \cite{karshenboim2005}, whereas the probability for inducing a $\pi^-$ 
transition is small.
Some transitions involve ultraviolet wavelengths, which are not easily accessible by high-power
lasers. The high-intensity $\pi^-$ beam ($>10^8$ s$^{-1}$) is
characterized by large momenta ($\ge 80$ MeV/c), momentum spread (5--10$\%$), and emittances. 
The contaminant electrons and $\mu^-$ in the beam, as well as the $\pi^-$ that immediately undergo 
nuclear absorption in the experimental apparatus, may give rise to backgrounds which prevent the 
detection of the $\pi{\rm He}^+$ laser resonance signal. The passage of charged particles in the helium target 
causes a broad spectra of scintillation photons emitted by helium excimers \cite{mckinsey2003}.
This paper mainly outlines the method by which the spectroscopic signal may be resolved rather than
the details of an apparatus which rejects these backgrounds.

This paper is organized in the following way. Section~\ref{history} reviews some past
research on $\pi{\rm He}^+$. In Sect.~\ref{estructure} we carry out three-body QED
calculations on the energy levels and fine structure of the 
$\pi{\rm ^4He}^+$ and $\pi{\rm ^3He}^+$ isotopes.
Section~\ref{initial_capture} discusses the formation and electromagnetic cascade of $\pi{\rm He}^+$.
We calculate the radiative and Auger rates and simulate the population evolutions of
the metastable states. Section~\ref{laser_excitation} describes the method to detect the laser resonance. 
The resonance profiles of some candidate transitions are
simulated. The energy distributions of the neutrons, protons, deuterons, and tritons that emerge following 
the nuclear absorption of $\pi^-$ are described in some detail. Section~\ref{simulations} presents
Monte Carlo simulations to roughly estimate the signal-to-background ratio of the laser resonance. 
Conclusions and a discussion concerning the determination of the 
$\pi^-$ mass are given in Sect.~\ref{conclude}.

\section{Brief history}
\label{history}

In two experiments carried out in the 1960s \cite{fetkovich1963,block1963,block1965},
$\pi^-$ were allowed to come to rest in liquid-helium bubble chambers. The readout photographs were scanned 
for any $\pi^-\rightarrow\mu^-+\overline{\nu}_{\mu}$ decay arising from $\pi^-$ at rest. Such 
events were expected to occur very rarely
since the $\pi^-$ stopped in other target materials are normally absorbed into the atomic nuclei within picoseconds; this
leaves no time for $\pi^-\rightarrow\mu^-+\overline{\nu}_{\mu}$ decay to occur with a lifetime of $\tau_{\pi}\sim 26$ ns. In the surprising
case of helium targets, however, an anomalously large fraction ($\sim 10^{-2}$) of the $\pi^-$ were found to 
decay in this way. This implied that the $\pi^-$ were trapped into atomic orbitals 
that retain long lifetimes against nuclear absorption \cite{fetkovich1963,block1963,block1965}. 
The average cascade time from atomic formation to absorption 
was estimated to be $\tau_c=200$--400 ps, based on the measured ratio between
$\pi^-\rightarrow\mu^-+\overline{\nu}_{\mu}$ decay and nuclear absorption events, which were compared with a simple model
describing the formation and deexcitation of the atom.  A third experiment that involved $\pi^-$ stopped in a 
diffusion chamber filled with $^3{\rm He}$ gas of pressure $p\sim 1.8\times 10^6$ Pa 
deduced a similar cascade time of $\tau_c=140\pm 70$ ps \cite{zaimidoroga1967}. 
Anomalous longevities were also detected for other negatively-charged particles $K^-$ \cite{block1965,fetkovich1970} 
and $\Sigma^-$ \cite{Comber1974,fetkovich1975} stopped in liquid helium.

Condo \cite{condo1964} attempted to explain these results by suggesting that some of the 
$\pi^-$ forms the three-body $\pi{\rm He}^+$ atom via the reaction
\begin{equation}
\pi^- + {\rm He} \rightarrow  \pi{\rm He}^++e^-.
\label{firstimp}
\end{equation}
The principal quantum number of the initially populated states was assumed to be distributed around the value                                                 
\begin{equation}                                                                                 
n\sim n_0=\sqrt{\frac{M^{\ast}}{m_e}}. 
\label{n0}                                                             
\end{equation}                                                                                   
Here $m_e$ and $M^{\ast}$ denote the electron mass and the reduced mass of the $\pi^-$--${\rm He}^{2+}$ 
pair, respectively. The $n_0\sim 16$ value for $\pi{\rm ^4He}^+$ and $\pi{\rm ^3He}^+$ 
corresponds to the $\pi^-$ orbitals with the same radius and binding energy as that 
of the displaced $1s$ electron in the reaction of Eq.~\ref{firstimp}. 
These Rydberg states are long-lived
\cite{day1960,borie1980,landua1982,jensen2002,cohen2004,raeisi2009,popov2012} since
(i): the wave functions of the $\pi^-$ orbitals have very little overlap with the 
nucleus and so the $\pi^-$ cannot be directly absorbed,
(ii): the states have long ($>10$ ns) lifetimes against radiative deexcitation of the $\pi^-$,
(iii): the deexcitation to a $\pi{\rm He}^{2+}$ ionic state of principal and angular momentum quantum
numbers of $n_i\sim 13$ and $\ell_i=n_i-1$ by Auger emission of the remaining electron
\begin{equation}
\pi\mathrm{He}^+_{(n,\ell)}\to \pi\mathrm{He}^{2+}_{(n_i,\ell_i)} + e^-
\end{equation}
is suppressed because of the large binding energy ($\sim 25$ eV) of the electron and the 
high multipolarity $\Delta\ell_A =\ell-\ell_i\ge 3$ of the Auger transition, and
(iv): with this electron in place, the $\pi^-$ is protected against Stark mixing during atomic collisions.

Russell \cite{russell1969,russell1970_1,russell1970_2,russell1970_3} calculated the Auger and 
radiative decay rates of 
the $\pi{\rm He}^+$ states. The Auger rates $A_{(15,14)}=2\times 10^{12}$ s$^{-1}$ and 
$A_{(16,15)}=4\times 10^9$ s$^{-1}$ of the states $(n,\ell)=(15,14)$ and $(16,15)$
appeared to be too large \cite{russell1969} 
to account for the observed longevity of $\pi^-$ in helium. Fetkovich \cite{fetkovich1971} 
suggested that some $\pi^-$ are capture into the $n>17$ states, which have smaller Auger rates. 
On the other hand, they noted inconsistencies \cite{fetkovich1971} between the experiments 
which deduced average cascade times of 200--400 ps in liquid-helium targets \cite{fetkovich1963,block1963},
and theoretical models of pionic atoms, which include collisional deexcitation processes that predicted a 
value of $\sim 20$ ps.

In 1989, an experiment at KEK directly measured the lifetime of $K^-$ stopped in liquid helium using particle 
counters \cite{yamazaki1989}. It showed that $\sim 98\%$ of the $K^-$ are 
promptly absorbed by the helium nucleus, whereas the remaining $(1.9\pm 0.3)\%$ retain a lifetime of 
$9.5\pm 0.3$ ns against the nuclear absorption and free decays of $K^-$. 
A similar experiment carried out at TRIUMF \cite{nakamura1992} showed that $(2.30\pm 0.07)\%$ of the
$\pi^-$ stopped in liquid helium retain a lifetime of $7.3\pm 0.1$ ns. These lifetimes are much longer 
than the average cascade times deduced by the above cloud chamber experiments. We are
unaware of another measurement on the metastability of $\pi^-$ in helium \cite{fetkovich1963,block1963,zaimidoroga1967,nakamura1992}. 
Some calculations on the nonrelativistic energies and decay rates of $\pi{\rm He}^+$
have been presented in Ref.~\cite{boyan}. The present paper will provide higher-precision values 
that include relativistic and QED corrections over a larger range of states.

Experimental and theoretical efforts in the last 20 years have concentrated on the 
antiprotonic helium ($\overline{p}{\rm He}^+\equiv\overline{p}+{\rm He}^{2+}+e^-$) atom, which 
retain mean lifetimes of 3--4 $\mu{\rm s}$ against antiproton annihilation in the helium nucleus \cite{iwasaki1991}. 
The transition frequencies of $\overline{p}{\rm He^+}$ were recently measured to 
a fractional precision of $(2.3-5)\times 10^{-9}$ \cite{mhori2011} in laser spectroscopy experiments
\cite{Mor94,mhori2001,mhori2003,mhori2006}. By comparing the results with 
three-body QED calculations \cite{korobov2008}, the antiproton-to-electron mass ratio was determined as 
$M_{\overline{p}}/m_e=1836.1526736(23)$ \cite{mhori2011}. 
Most of the calculated radiative and Auger rates \cite{kor97,korobov2008} agreed 
with the state lifetimes measured by laser spectroscopy 
\cite{mhori1998,*mhori1998e,mhori2001,yamaguchi2002,yamaguchi2004} within a 
precision of 10--30$\%$. A similar experiment should be possible for $\pi{\rm He}^+$.

\section{Energy level structure}
\label{estructure}

\subsection{Spin-independent part}
\label{spp}

As the $\pi{\rm He}^+$ states are unstable against Auger decay, they are properly described 
as pseudo- or resonant states which lie in the continuum of the non-relativistic three-body Hamiltonian, 
rather than truly bound states.
The resonances constitute poles of the scattering matrix in the complex momentum plane,
which in turn can be mapped to the unphysical sheets of the energy Riemann surface \cite{burke2011}. 
Direct numerical calculation of these eigenvalues is difficult since the stationary wavefunctions 
exponentially diverge when the distances between the constituent particles tend to infinity. 
To transform the wavefunctions into convergent bound-state forms 
that allows the $\pi{\rm He}^+$ energy levels to be readily calculated, we employ the complex-coordinate rotation (CCR) 
method \cite{Ho83,moiseyev1998}. The coordinates of the dynamical system are continued (``rotated") to the complex 
plane, using the transformation $r_{ij}\rightarrow r_{ij} e^{i\varphi}$, where $\varphi$ denotes a rotational angle. 
Under this transformation, the Hamiltonian changes as a function of $\varphi$,
\begin{equation}
\hat{H}_{\varphi} = \hat{T} e^{-2 i \varphi} + \hat{V} e^{-i \varphi},
\end{equation}
where $\hat{T}$ and $\hat{V}$ denote the kinetic energy and Coulomb potential operators \cite{Ho83,kino1998,korobov2003}. 
By this analytical transformation, the continuum spectrum of the original Hamiltonian is rotated around branch points
(or thresholds). The resonant poles corresponding to $\pi{\rm He}^+$ states are thus uncovered by branch cuts, 
so that they belong to the discrete spectrum of the rotated $\hat{H}_{\varphi}$.
The resonance energy can be determined by solving the complex eigenvalue problem for the rotated Hamiltonian,
\begin{equation}\label{roteqn}
(\hat{H}_{\varphi} - E)\Psi_{\varphi} = 0,
\end{equation}
as the eigenfunction $\Psi_{\varphi}$ is square-integrable. The complex eigenvalue 
$E = E_r - i\Gamma/2$ defines the energy $E_r$ and the width $\Gamma$ of the resonance; the latter is related to the Auger rate of the state by $A=\Gamma/\hbar$, where the reduced Planck constant is denoted by $\hbar$.

As we mentioned above, the $\pi{\rm He}^+$ states are conventionally  
characterized by the approximate quantum numbers $(n,\ell)$ of the atomic $\pi^-$ orbital \cite{condo1964,russell1969,russell1970_1,*russell1970_2,*russell1970_3}. The states also share the features of a 
polar molecule \cite{shimamura1992} with two nuclei ${\rm He}^{2+}$ and $\pi^-$, which are described 
by an alternative pair of quantum numbers $(v,L)$.  Here $v$ denotes the 
vibrational quantum number, or equivalently, the number of radial nodes of the $\pi^-$ wavefunction.
The total orbital angular momentum quantum number is denoted by $L$. As the electron occupies roughly 
the $1s$ state, the conversion between the two sets of quantum numbers follows,
\begin{eqnarray}
v&=&n-\ell-1, \\
L&=&\ell.
\end{eqnarray}
The $(v,L)$ numbers are used in our representation of the variational wavefunctions. 
All numerical results on the state energies, lifetimes, and populations are
presented using the conventional $(n,\ell)$ quantum numbers.

We utilize a variational wavefunction \cite{var99} which reflects both the atomic and molecular 
nature of $\pi{\rm He}^+$. The coordinate system is shown in Fig. 1 of Ref.~\cite{var99}.
The angular part of the wavefunction is described by the (molecular) bipolar harmonic expansion of the form
\begin{equation}\label{Schwexp}
\Psi^{L\Lambda}_M(\mathbf{R},\mathbf{r}) =
  \sum_{l+l_e=L} R^{l} r^{l_e}\{Y_{l} \otimes Y_{l_e} \}_{LM}G^{L\Lambda}_{ll_e}(R,r,\theta).
\end{equation}
The components $G^{L\Lambda}_{ll_e}(R,r,\theta)$ are functions of the internal degrees of freedom, 
which can be expanded in the (atomic) exponential form
\begin{equation}\label{variat}
 G^{L\Lambda}_{l l_e}(R,r,\theta) = \sum_{i=1}^{\infty} C_i
 e^{-\alpha_i R -\beta_i r -\gamma_i | \mathbf{R} - \mathbf{r} | }.
 \end{equation}
Here $M$ denotes the projection of the total orbital angular momentum on the $z$
axis of the fixed frame, and $\Lambda=(-1)^L$ the total spatial parity. 
 The complex parameters 
$\alpha_i, \beta_i$ and $\gamma_i$ are generated in a quasirandom way.

We used variational basis sets which includes 2000 functions. 
Tables~\ref{nonrela} and \ref{nonrela2} present the 
nonrelativistic energies and widths of the $\pi{\rm ^4He}^+$ and $\pi{\rm ^3He}^+$ states, respectively. 
The expectation values of the operators needed to evaluate the leading-order relativistic correction
for the bound electron and the one-loop self-energy and vacuum polarization corrections 
of order $R_{\infty}\alpha^3$ in atomic units (see Appendix)
are also shown. Here $R_{\infty}$ and $\alpha$ denote the Rydberg and fine structure constants,
respectively. The numerical precision of the nonrelativistic energies (indicated in parenthesis)
are better than $\sim 10^{-8}$, but the actual precision is limited to 
$>10^{-6}$ by the experimental uncertainty on the $\pi^-$ mass 
\cite{pdg} used in these calculations,
\begin{equation}
M_{\pi^-}=139.57018(35) {\rm MeV}/c^2.
\end{equation}
The uncertainties on the Auger widths are of similar magnitude to the nonrelativistic energies 
since these CCR calculations evaluate the ``complex energy"  of Eq.~\ref{roteqn}. 

\begin{figure*}[tb]
\begin{center}
\includegraphics[height=12cm]{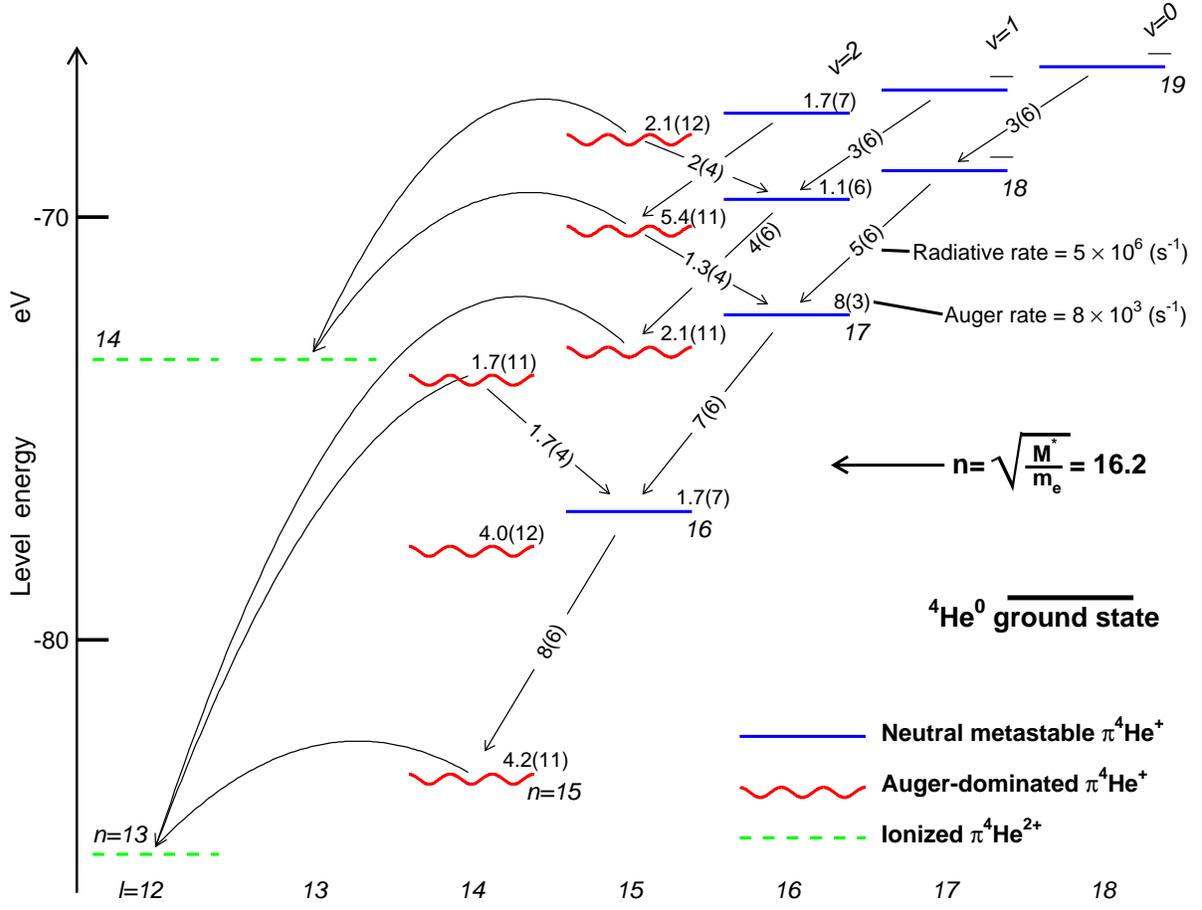}
\begin{minipage}[t]{16.5 cm}
\caption{\label{fig:pi_energy4} 
(Online color) Energy level diagram of $\pi{\rm ^4He}^+$ atom. 
On the left-hand scale the theoretical absolute energy of each 
state $(n,\ell )$ is plotted relative to the three-body-breakup threshold.
The wavy lines indicate Auger-dominated states with picosecond-scale lifetimes,
solid lines the metastable levels with lifetimes of $>10$ ns.
The Auger decay rate of each state is shown in s$^{-1}$.
The dashed lines indicate the final 
$\pi{\rm He}^{2+}$ ionic states formed after Auger electron emission, the curved arrows
some Auger transitions with minimum $|\Delta\ell_A|$. Radiative transitions of the 
types $(n,\ell)\rightarrow (n-1,\ell-1)$ and $(n,\ell)\rightarrow (n-1,\ell+1)$ are 
indicated using straight arrows, with the corresponding decay rates shown in s$^{-1}$. 
}
\end{minipage}
\end{center}
\end{figure*}

\begin{figure*}[tb]
\begin{center}
\includegraphics[height=12cm]{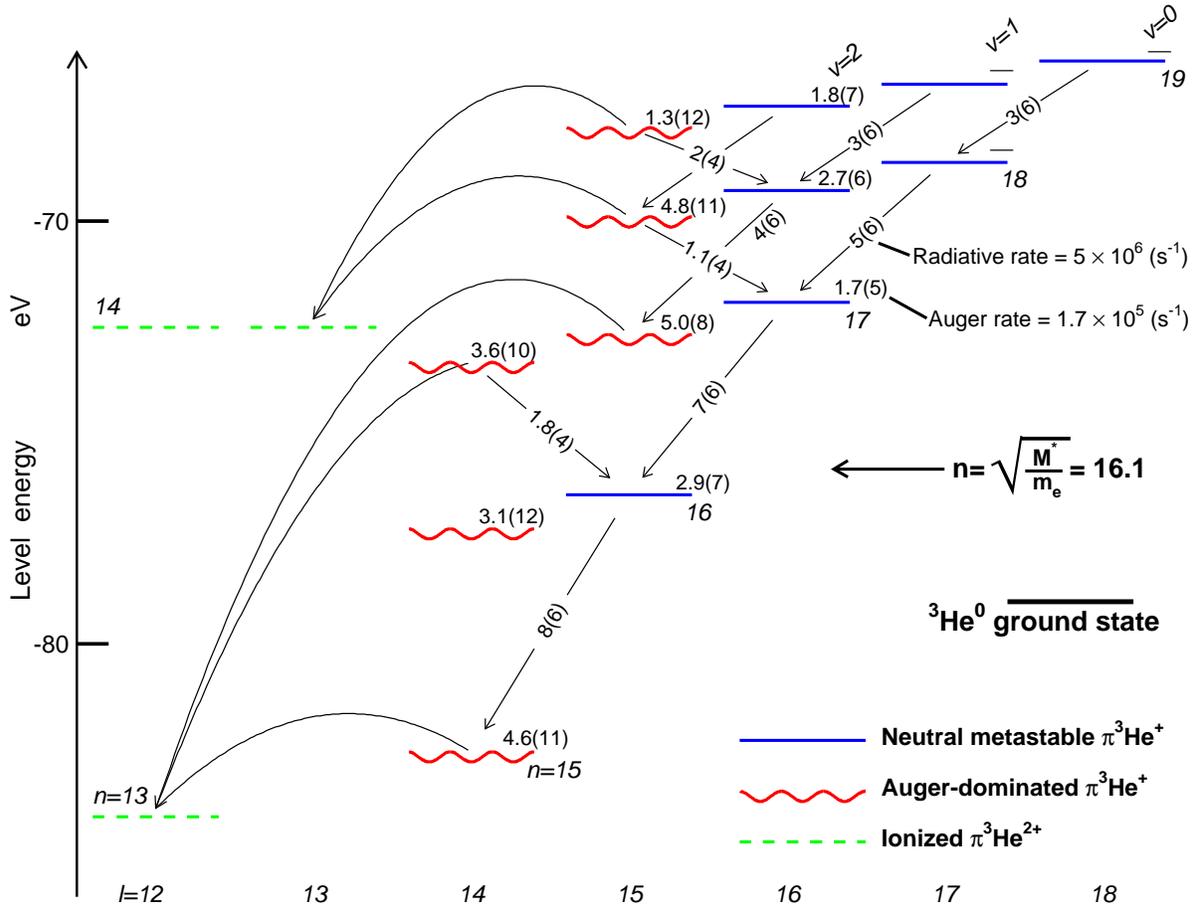}
\begin{minipage}[t]{16.5 cm}
\caption{\label{fig:pi_energy3} 
(Online color) Energy level diagrams of $\pi{\rm ^3He}^+$ atom and $\pi{\rm ^3He}^{2+}$ ion. 
See caption of Fig.~\ref{fig:pi_energy4} for details.}
\end{minipage}
\end{center}
\end{figure*}

The energy-level diagrams of $\pi{\rm ^4He}^+$ and $\pi{\rm ^3He}^+$ 
in the $n=15$--19 regions are shown in Figs.~\ref{fig:pi_energy4} and 
\ref{fig:pi_energy3}. The level energies relative to the three-body breakup threshold
are indicated in solid or wavy lines. Radiative transitions of the type $(n,\ell)\rightarrow (n-1,\ell-1)$
involve energy intervals of $\Delta E=2-6$ eV, whereas the 
energies of states with the same $n$-value increase in steps of $\Delta E=0.5-0.7$ eV 
for every change $\Delta\ell=+1$. This removal of the $\ell$-degeneracy
suppresses Stark mixing during atomic collisions.

The energy levels of the two-body $\pi{\rm ^4He}^{2+}$ 
and $\pi{\rm ^3He}^{2+}$ ions in the regions $n_i=13$--14 are shown by dashed lines, 
superimposed on the same figures. 
The spin-independent parts of the $\pi{\rm He}^{2+}$ level energies can be calculated to a fractional precision 
better than $10^{-5}$ using the simple Bohr formula in atomic units,
\begin{equation}
E_{n_i}=-\frac{2M^{\ast}}{n_i^2},
\end{equation}
where $M^{\ast}$ denotes the reduced mass of the $\pi^-$--${\rm He^{2+}}$ pair.
Radiative transitions of the type $(n_i,\ell_i)\rightarrow (n_i-1, \ell_i\pm 1)$ lie in 
the ultraviolet ($>10$ eV) region. The $\pi^-$ states with the same $n_i$-value are now degenerate, so that
Stark mixing with $s$, $p$, and $d$ states occurs during collisions with helium atoms
\cite{day1960,borie1980,landua1982}. This normally leads to $\pi^-$ absorption within picoseconds, 
although lifetimes of $\sim 100$ ns
have been observed for $\overline{p}{\rm He}^{2+}$ ions formed in low-density targets where
the collision rate is sufficiently low \cite{mhori2005}. The short lifetime and large transition 
energies make it difficult to induce laser resonances in $\pi{\rm He}^{2+}$ and so we will not 
consider the feasibility here.

\subsection{Fine structure}

The boson-boson $\pi^-$--${\rm ^4He}^{2+}$ pair has no spin-spin or spin-orbit interactions. 
The coupling $\mathbf{s}_e\cdot \mathbf{L}$ between the spin vector $\mathbf{s}_e$ 
of the 1s electron and the orbital angular momentum vector $\mathbf{L}$ of the $\pi^-$ splits each 
$\pi{\rm ^4He}^+$ state $(n,\ell)$ into a pair of fine structure substates, which are characterized
by the total angular momentum vector
\begin{equation}
\mathbf{J}=\mathbf{L}+\mathbf{s}_e.
\end{equation}
This fine structure splitting is determined by the effective Hamiltonian using the corresponding operators
\begin{equation}
\hat{H}_{\rm eff} = E_1\cdot \left(\mathbf{\hat{s}}_e\cdot\mathbf{\hat{L}}\right),
\label{heff}
\end{equation}
where the energies $E_1$ (shown in Tables~\ref{nonrela} and \ref{nonrela2}) are calculated 
by integrating over the spatial internal degrees of freedom. The expectation value of the
scalar product in Eq.~\ref{heff} may be expanded using the total angular momentum 
quantum number $J$ as
\begin{eqnarray}
\langle\mathbf{\hat{s}}_e\cdot\mathbf{\hat{L}}\rangle&=&
  \frac{1}{2}\left[
     \langle\mathbf{\hat{J}}^2\rangle-\langle\mathbf{\hat{L}}^2\rangle-\langle\mathbf{\hat{s}}_e^2\rangle
  \right] \nonumber \\
  &=& \frac{1}{2}\left[
     J(J+1)-L(L+1)-3/4
  \right].
\end{eqnarray}

\begin{figure*}[tb]
\begin{center}
\includegraphics[height=11cm]{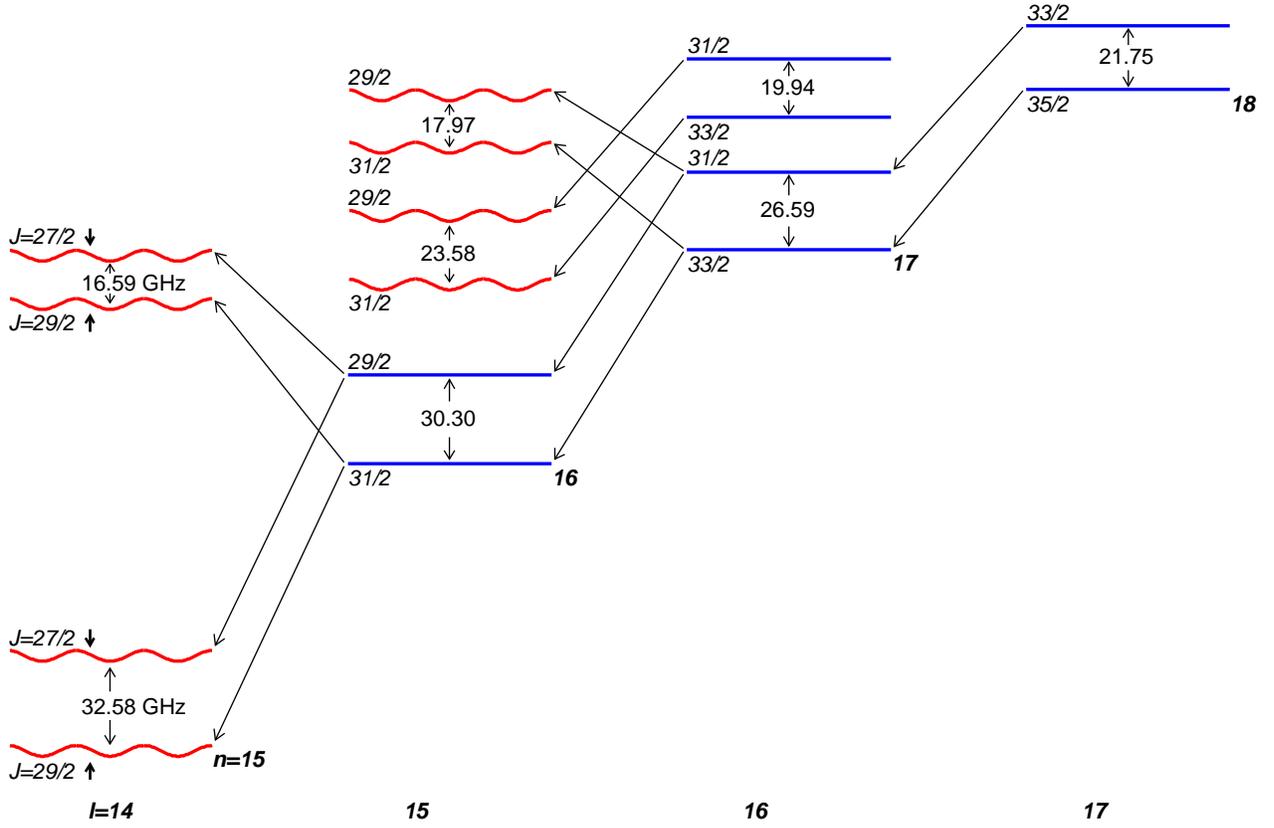}
\begin{minipage}[t]{17 cm}
\caption{\label{hfspihe4} 
(Online color) Fine structure of $\pi{\rm ^4He}^+$, indicating the principal-, orbital angular momentum-,
and total angular momentum quantum numbers $n$, $\ell$, and $J$ of each substate. Wavy lines show
Auger-dominated short-lived states with picosecond-scale lifetimes, solid lines metastable states with
$>10$ ns lifetimes. Small bold arrows indicate the direction of the electron spin. Fine structure 
splittings $\Delta\nu_{\rm fs}$ of each substate pairs shown in gigahertz. Some $E1$ laser transitions 
that do not flip the electron spin are indicated by diagonal arrows.  Drawing not to scale.}
\end{minipage}
\end{center}
\end{figure*}

In the $n=16$--18 region (Fig.~\ref{hfspihe4}), the fine structure splitting increases from 
$\Delta\nu_{\rm fs}=16.6$ to 32.6 GHz for  $\pi{\rm ^4He}^+$ states with smaller 
$n$- and larger $\ell$-values. Each resonance profile for a transition 
of the type $(n,\ell)\rightarrow(n-1,\ell-1)$ or $(n,\ell)\rightarrow(n+1,\ell-1)$
contains a dominant pair of fine structure sublines separated by 2--5 GHz or
4--9 GHz, respectively. These sublines correspond to the pairs of $\Delta J=-1$ transitions
(indicated by arrows in Fig.~\ref{hfspihe4}) that do not flip the electron spin. 
They cannot be resolved as distinct peaks for transitions that involve 
states with large Auger widths ($A_{(n,\ell)}/2\pi\ge 25$ GHz, wavy lines).
The laser resonance profile also contains a weaker (by $\sim 10^{-2}$) subline, 
which corresponds to the spin-flip transition $\Delta J=0$. 

In the $\pi{\rm ^3He}^+$ case, a hyperfine 
structure arises from the spin-spin interaction between the electron and $^3$He nucleus,
as in $\overline{p}{\rm ^3He}^+$ \cite{mhori2006,friedreich2013}.
The size of this hyperfine splitting is $<10^{-1}$ of the fine splitting $\Delta\nu_{\rm fs}$.

\section{Atomic cascade}
\label{initial_capture}

\subsection{Primary populations}
\label{popu}

Only the $\pi{\rm He}^+$ transitions that involve states with sizable $\pi^-$ populations can be
detected by laser spectroscopy. Nothing is experimentally known about the $(n,\ell)$ quantum 
numbers of the primordial (i.e., initially occupied) states after the formation process of Eq.~\ref{firstimp}. 
It is difficult to predict the populations by modeling the processes by which the 
$\pi^-$ slows down and is captured by a helium atom. This is partially due to the large number of couplings 
between the initial $\pi^-$ continuum and bound final $\pi{\rm He}^+$ states. The simple model of 
Eq.~\ref{n0} \cite{baker1960,day1960,landua1982} predicts capture into the $n\sim 16$ region. 
The newly-formed $\pi{\rm He}^+$ recoils with roughly the same 
momentum as the incoming $\pi^-$. By energy conservation, its binding energy is equal to
\begin{equation}
B_n=I_0-\frac{T_{\pi}M_{\rm He}}{M_{\rm He}+M_{\pi^-}}+T_e.
\label{n1}
\end{equation}
Here $I_0\sim 24.6$ eV denotes the ionization potential of helium, $T_{\pi}$ and $T_e$ are
the laboratory energies of the incoming $\pi^-$ and ejected electron, respectively, 
and $M_{\rm He}$ is the mass of the helium nucleus. 

The $\overline{p}{\rm ^4He}^+$ and $\overline{p}{\rm ^3He}^+$ isotopes are the only 
exotic atoms for which the primary populations have been experimentally studied 
so far. Reference~\cite{mhori2002} measured the intensities of some laser-induced resonances,
which are proportional to the number of antiprotons occupying the parent state of the transition. 
In the $\overline{p}{\rm ^4He}^+$ case, the 
region $n=37$--40 was found to account for nearly all the observed metastability.
The largest population was observed for $n\sim 38$. 
The $\overline{p}{\rm ^3He}^+$ were distributed over $n=35$--38, with the
maximum at $37$. Very little population was inferred for states $n\ge 41$. 
These results appear to support the predictions $n_0=38.3$ and $37.1$ of Eq.~\ref{n0} for 
the $\overline{p}{\rm ^4He}^+$ and $\overline{p}{\rm ^3He}^+$ cases.

We assume that the $\pi{\rm He}^+$ populations are distributed 
over the $n=15$--17 states, which have the same binding energies $B_n$
as the populated $\overline{p}{\rm He}^+$ states in the above experiments.
The Auger rate calculations of Sect.~\ref{augerrates} show that only two $\pi{\rm He}^+$
states in this range are long-lived. One of these $(16,15)$
is expected to contain the largest population according to Eq.~\ref{n0}. The 
population in $(17,16)$ may be smaller, whereas the $n\ge 18$ states retain negligible
metastable population. These assumptions restrict the candidate transitions that 
can be studied by laser spectroscopy. 

Before proceeding, we note that many theoretical models 
\cite{cohen1983,ohtsukiprivat,dolinov1989,korenman1996a,beck1993,briggs1999,cohen2000,tokesi2005,revai2006,tong2008,sakimoto2010} on the formation of $\overline{p}{\rm He}^+$ indicate 
that capture can occur into states with $n$-values higher than those implied 
by Eq.~\ref{n0}. These models claim that the electron in Eq.~\ref{n1} is ejected with nearly zero energy,
i.e., $T_e\ll I_0$. Only low-energy $T_{{\overline{p}}}\ll I_0$ antiprotons are
captured into the $n_0\sim 38$ states with binding energies $B_n\sim I_0$, whereas                                            
$T_{\pi}\sim 25$-eV antiprotons are captured into much higher $B_n\sim 0$ regions. 
Some of these calculations predict that 
$12-25\%$ of the antiprotons stopped in helium will be captured into 
metastable states, primarily in the $n\ge 41$ region. This seems to contradict
the experimental results \cite{mhori2002} that show little population in $n\ge 41$; 
in fact, the models overestimate the fraction of antiprotons occupying the metastable 
states by an order of magnitude. The reason for this is not understood.
Korenman \cite{korenman1996b} suggested that the $\overline{p}{\rm He}^+$                                   
formed in the high-$n$ states recoil with large kinetic energies, and are rapidly 
destroyed by collisions. Sauge and coworkers \cite{sauge2001} pointed out
that the quenching cross-sections for these states may be large even at thermal                                
energies. It has also been suggested that the antiprotons captured into the 
high-$n$ states have relatively small $\ell$-values, so that they quickly deexcite radiatively                                      
\cite{cohen2000}. Similar effects in the $\pi{\rm He}^+$ case have not been 
theoretically studied so far.

\subsection{Radiative decay rates}

The slow radiative deexcitation of the $\pi^-$ by emitting visible or UV photons is 
not a dominant cascade mechanism in $\pi{\rm He}^+$ compared to
the faster process of $\pi^-\rightarrow\mu^-+\overline{\nu}_{\mu}$ decay.
This is in contrast to the $\overline{p}{\rm He}^+$ case, in which an 
antiproton can radiatively cascade through several metastable states.

Table~\ref{transitions} shows the reduced matrix elements 
$\langle L'v'\|\mathbf{\hat{d}}\|Lv \rangle$ of the dipole electric moment operator
\begin{equation}
\mathbf{\hat{d}}=\sum_{i=1}^3 Z_i\mathbf{\hat{R}}_i\,
\end{equation}
between some $\pi{\rm He}^+$ states. 
The charges and position operators of the three constituent particles in the center-of-mass frame are denoted by 
$Z_i$ and $\mathbf{\hat{R}}_i$. The values were calculated using the variational basis sets with
1000 functions. The corresponding radiative transition rates are obtained in atomic units as
\begin{equation}
\lambda=\frac{4}{3}\alpha^3\>(\Delta E)^3\>\frac{\langle L^{\prime}v^{\prime}\|\mathbf{\hat{d}}\|Lv \rangle^2}{2L+1}\,,
\end{equation}
where the transition energy is denoted by $\Delta E$. The values can be converted to SI units using the factor of 1 a.u. (of time) equal to $2.4189\cdot10^{-17}$ s.

The results indicate that radiative deexcitation preferentially proceeds via 
the type $(n,\ell)\rightarrow (n-1,\ell-1)$
that keeps the radial node number $v$ constant. The decay 
rates of these favored transitions of $\pi{\rm ^4He}^+$ and $\pi{\rm ^3He}^+$
are indicated in Figs.~\ref{fig:pi_energy4} and \ref{fig:pi_energy3}. They range 
from $\sim 3\times 10^6$ s$^{-1}$ for the transition 
$(19,18)$$\rightarrow$$(18,17)$, to
$\sim 8\times 10^6$ s$^{-1}$ for $(16,15)$$\rightarrow$$(15,14)$ in both isotopes. 

The radiative decay rates of unfavored transitions of the type $(n,\ell)\rightarrow (n-1,\ell +1)$ 
were also calculated. The rates $(1-2)\times 10^{4}$ s$^{-1}$ for 
$(19,15)$$\rightarrow$$(18,16)$,
$(18,15)$$\rightarrow$$(17,16)$, and
$(17,14)$$\rightarrow$$(16,15)$ were two orders of magnitude
smaller than those of the favored transitions. This is because of the small spatial 
overlap between the wave functions of the parent and daughter states with 
different ($\Delta v=-2$) radial node numbers. 

\subsection{Auger decay rates}
\label{augerrates}

The $\pi{\rm ^4He}^+$ and $\pi{\rm ^3He}^+$ states $(n,\ell)$ indicated by wavy lines in 
Figs.~\ref{fig:pi_energy4} and \ref{fig:pi_energy3}, respectively, have large Auger decay rates 
$A_{(n,\ell)}=10^{10}-10^{12}$ s$^{-1}$. 
Most of them are connected to energetically lower-lying $\pi{\rm He}^{2+}$ ionic 
states (dashed lines) via Auger decays with small multipolarities $\Delta\ell_A\le 2$.  
On the other hand, the decays from $\pi{\rm He}^{+}$ states 
(solid lines) with multipolarities $\Delta\ell_A\ge 3$ are slow, i.e., $A_{(n,\ell)}<10^{8}$ s$^{-1}$. 
The $\pi{\rm He}^+$ states are therefore grouped into two regions: 
the metastable states of $\Delta\ell_A\ge 3$ dominated by $\pi^-\rightarrow e^-+\overline{\nu}_{\mu}$
decay, and the Auger-dominated short-lived states of 
$\Delta\ell_A\le 2$. An exception to this rule is the large rate $A_{(17,15)}\sim 5\times 10^{8}$ s$^{-1}$
of the $\overline{p}{\rm ^3He}^+$ state $(17,15)$, which lies slightly below the energy 
needed to make a $\Delta\ell_A=2$ transition to the ionic state $(n_i,\ell_i)=(14,13)$.

The Auger rates of the $\pi{\rm ^4He}^+$ states $(n,\ell)=(15,14)$ and $(16,15)$ calculated by us
are 10--100 times smaller than those of Ref.~\cite{russell1969}. The state $(16,15)$ which 
contains the largest population according to Eq.~\ref{n0} is predicted to have a 60-ns lifetime 
against Auger decay. This state is therefore a prime candidate for laser spectroscopy.

\subsection{Cascade model}                                                                                                 
\label{cascsect}

\begin{figure}[tb]
\begin{center}
\includegraphics[height=5cm]{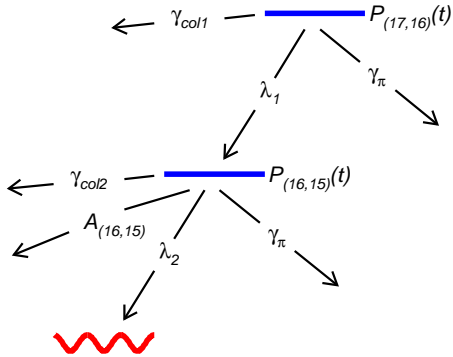}
\begin{minipage}[t]{8.5 cm}
\caption{\label{twolevel} 
(Online color) Schematic drawing of the cascade model involving two states
$(n,\ell)=(16,15)$ and $(17,16)$ of metastable $\pi{\rm He}^+$.
The state populations are denoted by $P_{(n,\ell)}(t)$,
the Auger decay rates by $A_{(n,\ell)}$, and the collisional 
deexcitation rates by $\gamma_{\rm col1}$ and $\gamma_{\rm col2}$.
The radiative rates of the transitions 
$(17,16)\rightarrow (16,15)$ and
$(16,15)\rightarrow (15,14)$ are denoted by $\lambda_1$ and $\lambda_2$,
and the $\pi^-\rightarrow\mu^-+\overline{\nu}_{\mu}$ decay rate by
$\gamma_{\pi}=3.8\times 10^7$ s$^{-1}$.}
\end{minipage}
\end{center}
\end{figure}

The time evolutions of the populations $P_{(n,\ell)}(t)$ in the metastable $\pi{\rm ^4He}^+$ states are
simulated using the cascade model depicted in Fig.~\ref{twolevel}. A $\sim 2.3\%$ fraction of the $\pi^-$ that 
comes to rest in a liquid helium target are long-lived \cite{nakamura1992}; we assume that these 
are captured into the states $(16,15)$ and $(17,16)$ (Sec.~\ref{popu}).
The remaining $\sim 98\%$ of the $\pi^-$ are promptly absorbed 
into the helium nuclei. The metastable populations evolve as
\begin{eqnarray}
dP_{(17,16)}=&-&\large(\lambda_{1}+\gamma_{\pi}+\gamma_{\rm col1}\large)P_{(17,16)}dt, \nonumber \\
dP_{(16,15)}=&-&\large(\lambda_{2}+A_{(16,15)}+\gamma_{\pi}+\gamma_{\rm col2}\large)P_{(16,15)}dt \nonumber \\
                          &+&\lambda_{1}P_{(17,16)}dt.
\label{pop}
\end{eqnarray}
Here $\lambda_{1}$ and $\lambda_2$ denote the radiative decay rates of the transitions
$(17,16)\rightarrow (16,15)$ and $(16,15)\rightarrow (15,14)$, and
$\gamma_{\pi}=3.8\times 10^7$ s$^{-1}$ the $\pi^-\rightarrow\mu^-+\overline{\nu}_{\mu}$ decay rate.
The small Auger decay rate $A_{(17,16)}\sim 8\times 10^{3}$ s$^{-1}$ of $(17,16)$ is neglected. 
We assume that this model includes the important cascade mechanisms which affect the metastable 
populations. For example, some of the $\pi{\rm ^4He}^+$ that populate $(16,15)$ can radiatively 
decay to $(15,14)$ at a rate $\lambda_2=8\times 10^6$ s$^{-1}$. These atoms no longer retain 
their metastability since they undergo Auger decay and $\pi^-$ absorption within picoseconds. 

Collisions between $\pi{\rm ^4He}^+$ and helium atoms may cause 
other types of transitions, but the rates cannot be accurately predicted
because of the complexities of the reactions \cite{korenman1996b,sauge2001,sakimoto2012}. 
The rates of collisional deexcitations that destroy the populations in $(17,16)$ and $(16,15)$ 
are denoted by $\gamma_{\rm col1}$ and 
$\gamma_{\rm col2}$ in Eq.~\ref{pop}. Collisional shortening of some state lifetimes have 
been observed in $\overline{p}{\rm He}^+$ \cite{mhori1998,*mhori1998e,mhori2004}.

\begin{figure}[htbp]
\begin{center}
\includegraphics[height=8cm]{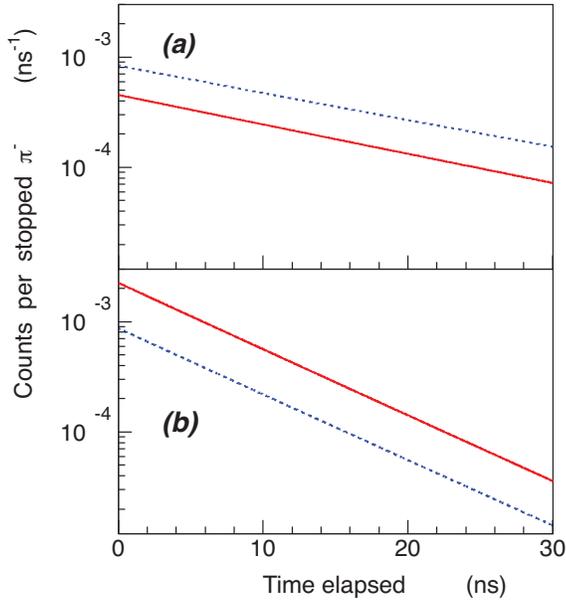}
\begin{minipage}[t]{8.5 cm}
\caption{\label{cascade_spectrum} 
(Online color) Time evolutions of the total count rates of metastable $\pi{\rm ^4He}^+$ that undergo 
$\pi^-$ nuclear absorption (solid lines) and $\pi^-\rightarrow\mu^-+\overline{\nu}_{\mu}$ decay (dotted lines), 
simulated for collisional deexcitation rates of (a) $\gamma_{\rm col1}=\gamma_{\rm col2}=0$, 
and (b) $8\times 10^7$ s$^{-1}$. The count rates are normalized to the total number of $\pi^-$
that come to rest in the helium target. Prompt nuclear absorptions of $\pi^-$ that occur at $t=0$ are not shown.}
\end{minipage}
\end{center}
\end{figure}

The normalized count rates of the metastable $\pi{\rm ^4He}^+$ that undergo nuclear absorption and $\pi^-\rightarrow\mu^-+\overline{\nu}_{\mu}$ 
decay can be calculated, respectively, as
 \begin{eqnarray}
\Gamma_{\rm abs}(t)&=&\frac{\gamma_{\rm col1}}{\lambda_{1}+\gamma_{\pi}+\gamma_{\rm col1}}\frac{dP_{(17,16)}}{dt} \nonumber \\
&+&\frac{\lambda_{2}+A_{(16,15)}+\gamma_{\rm col2}}{\lambda_{2}+A_{(16,15)}+\gamma_{\pi}+\gamma_{\rm col2}}\frac{dP_{(16,15)}}{dt}, \\
\Gamma_{\rm \pi\rightarrow\mu}(t)&=&\frac{\gamma_{\pi}}{\lambda_{1}+\gamma_{\pi}+\gamma_{\rm col1}}\frac{dP_{(17,16)}}{dt} \nonumber \\
   &+&\frac{\gamma_{\pi}}{\lambda_{2}+A_{(16,15)}+\gamma_{\pi}+\gamma_{\rm col2}}\frac{dP_{(16,15)}}{dt}. 
\label{abso}
\end{eqnarray}
In Fig.~\ref{cascade_spectrum} (a), the evolutions of $\Gamma_{\rm abs}(t)$ and $\Gamma_{\pi\rightarrow\mu}(t)$ 
are plotted for times $t=0-30$ ns in solid and dotted lines. The Auger and radiative decay rates 
are fixed to the theoretical values, while collisional deexcitations are neglected
($\gamma_{\rm col1}=\gamma_{\rm col2}=0$). 
The states $(17,16)$ and $(16,15)$ contain primary populations of $P_{(17,16)}(t=0)=0.005$ 
and $P_{(16,15)}(t=0)=0.018$, normalized to the total number of $\pi^-$ stopped in the helium target. 
The results of Fig.~\ref{cascade_spectrum} (a) show that integrated 
populations of $\int_0^{\infty}\Gamma_{\rm abs}dt=0.007$ and 
$\int_0^{\infty}\Gamma_{\rm \pi\rightarrow\mu}dt=0.016$ respectively undergo nuclear absorption and 
$\pi^-\rightarrow\mu^-+\overline{\nu}_{\mu}$ decay, with mean cascade 
lifetimes of $\tau_c=16-18$ ns .

This disagrees with the results of experiments \cite{nakamura1992} carried out using liquid-helium targets, 
which implies values of $\int_0^{\infty}\Gamma_{\rm abs}dt=0.017$,
$\int_0^{\infty}\Gamma_{\rm \pi\rightarrow\mu}dt=0.006$, and $\tau_c\sim 7$ ns, i.e., 
most of the long-lived $\pi^-$ undergo nuclear absorption instead of 
$\pi^-\rightarrow\mu^-+\overline{\nu}_{\mu}$ decay. The 7-ns decay lifetime of the experimental spectra in 
Ref.~\cite{nakamura1992} cannot be reproduced by adjusting the primary populations 
$P_{(n,\ell)}(t=0)$ alone in our model, as the calculated decay rates in Fig.~\ref{fig:pi_energy4} appear to be too small.

The spectrum of Fig.~\ref{cascade_spectrum} (b) was obtained by assuming collisional
deexcitation rates of
$\gamma_{\rm col1}\sim\gamma_{\rm col2}\sim 8\times 10^7$ s$^{-1}$. The other 
parameters are the same as those used in Fig.~\ref{cascade_spectrum} (a). The mean cascade lifetime 
$\tau_c\sim 7$ ns and the populations $\int_0^{\infty}\Gamma_{\rm abs}dt=0.017$ and 
$\int_0^{\infty}\Gamma_{\rm \pi\rightarrow\mu}dt=0.006$ destroyed via the two channels now 
agree with the experimental results. This model is used in our simulation of the laser spectroscopic signal in the following sections.
 
\section{Laser spectroscopic method}
\label{laser_excitation}

\subsection{Laser transitions}
 \label{tranprob}

\begin{figure*}[htbp]
\begin{center}
\includegraphics[height=10.8cm]{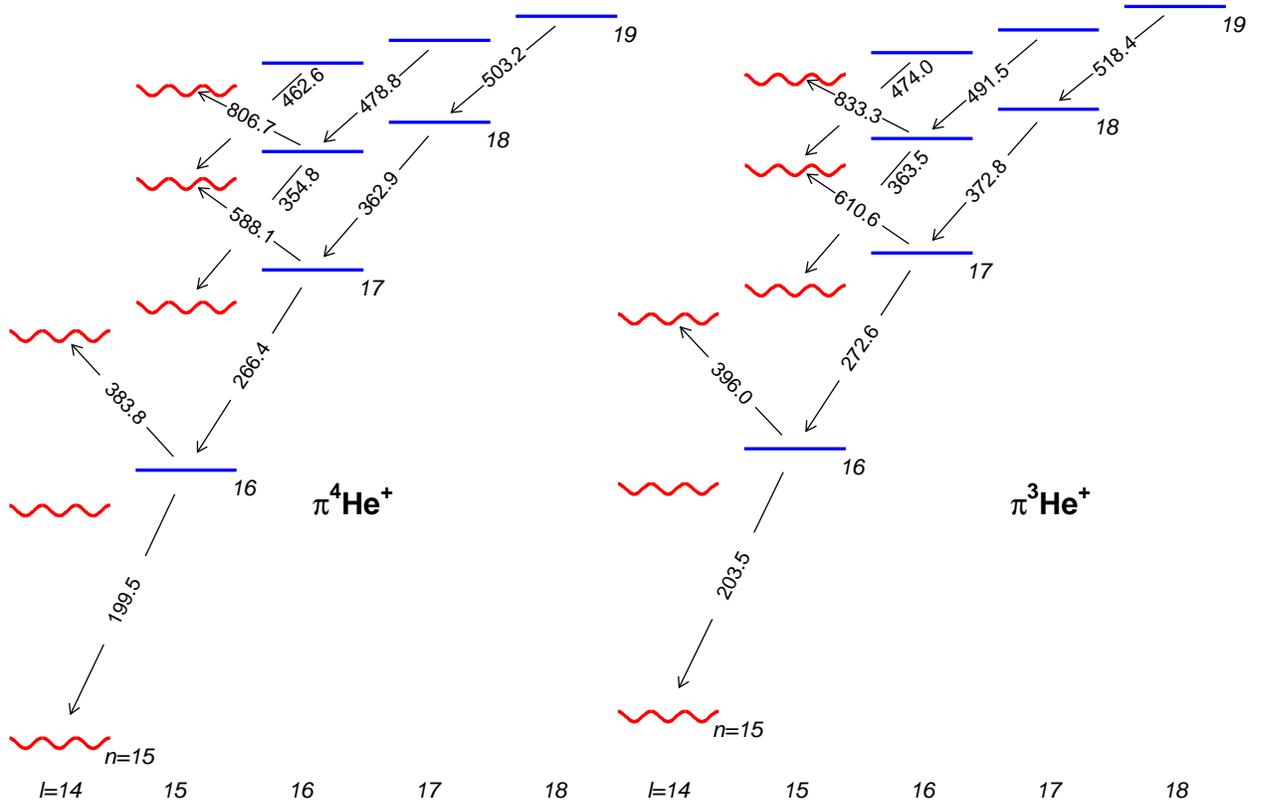}
\begin{minipage}[t]{16.5 cm}
\caption{\label{wavepihe4} 
(Online color) Energy level diagram of $\pi{\rm ^4He}^+$ and $\pi{\rm ^3He}^+$ isotopes.
Arrows indicate the laser transitions of the types $(n,\ell)\rightarrow (n-1,\ell-1)$ and
$(n,\ell)\rightarrow (n+1,\ell-1)$. The transition wavelengths are shown in nanometers.}
\end{minipage}
\end{center}
\end{figure*}

As mentioned earlier, we will excite laser transitions from the metastable 
$\pi{\rm He}^+$ states (indicated in Fig.~\ref{wavepihe4} by solid lines) to the Auger-dominated 
short-lived states (wavy lines). The resonance condition between the laser and $\pi{\rm He}^+$ is
revealed as a sharp spike in the rate of neutrons, protons, deuterons, and tritons that emerge 
(Sect.~\ref{absorption}) from the resulting $\pi^-$ absorption.

Fig.~\ref{wavepihe4} and Table~\ref{transitions} show some $E1$ transition wavelengths and frequencies
which include QED corrections. The strongest resonance signals are expected for the transitions 
$(n,\ell)=(16,15)$$\rightarrow$$(15,14)$, $(16,15)$$\rightarrow$$(17,14)$, 
and $(17,16)$$\rightarrow$$(18,15)$ in $\pi{\rm ^4He^+}$, since the parent states presumably  
contain large populations (Sec.~\ref{popu}). The favored transition $(16,15)$$\rightarrow$$(15,14)$ has 
the largest transition amplitude, but laser light at the resonant wavelength of $199.5$ nm is difficult to generate. 
Laser beams of $383.8$ nm and 588.1 nm for the transitions $(16,15)$$\rightarrow$$(17,14)$ 
and $(17,16)$$\rightarrow$$(18,15)$ can be readily produced using Ti:sapphire or dye lasers. 

The laser fluence $I$ needed to excite these transitions within the 7-ns lifetime of $\pi{\rm ^4He}^+$ 
is roughly estimated in the following way. The transition matrix element of the single-photon transition 
$(v,L,J,M_J)\rightarrow (v',L',J',M_J)$ from a substate of vibrational, orbital angular momentum, total 
angular momentum, and total azimuthal quantum numbers $v$, $L$, $J$, and $M_J$ induced by 
linearly-polarized laser light can be calculated using the Wigner 3-j and 6-j symbols as,
\begin{eqnarray}
\kappa_{JJ'M_J}&=&
(-1)^{J-M_J}
\left(
\begin{array}{@{\,}ccc@{\,}}
J      & 1 & J' \\
-M_J     & 0 & M_J \\
\end{array}
\right) \nonumber \\
&\times&(-1)^{J+L'-\frac{1}{2}}\sqrt{(2J+1)(2J'+1)} 
\left\{
\begin{array}{@{\,}ccc@{\,}}
L'      & J' & \frac{1}{2} \\
J     & L & 1 \\
\end{array}
\right\}
\nonumber \\
&\times &\langle L'v'\|\mathbf{\hat{d}}\|Lv \rangle .
\end{eqnarray}
This is related to the Rabi oscillation frequency $\Omega_{JJ'M_J}/2\pi$ in atomic units induced 
by a linearly-polarized, resonant laser field of amplitude $F$ by,
\begin{equation}
\frac{\Omega_{JJ'M_J}({\rm a.u.})}{2\pi}=\left|\kappa_{JJ'M_J}\right| F
\end{equation}
The $\pi{\rm ^4He}^+$ primary populations
are assumed to be uniformly distributed over the $\sim 60$ fine-structure and magnetic substates of the resonance parent state. 
The optical Bloch equations \cite{mhori2004,mhori2010} between pairs of states 
$(v,L,J,M_J)$ and $(v',L',J',M_J)$ are then numerically integrated, and the results averaged over all substates. 

\begin{figure}[tb]
\begin{center}
\includegraphics[height=10.5cm]{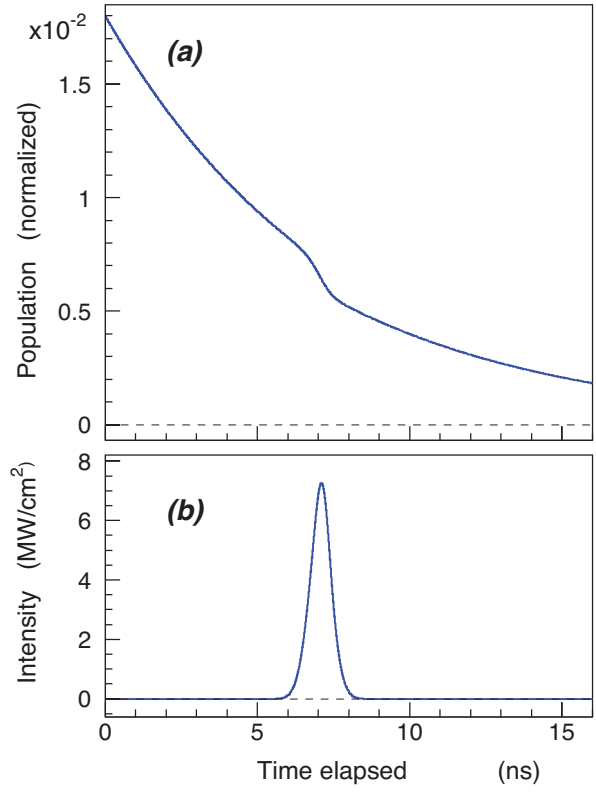}
\begin{minipage}[t]{8.5 cm}
\caption{\label{lasertrans} 
(Online color) (a) Population evolution of the $\pi{\rm ^4He}^+$ state $(n,\ell)=(16,15)$ as
a function of time elapsed. At $t=7$ ns, a 1-ns-long laser pulse excites the transition 
$(16,15)\rightarrow (17,14)$. This resulting in a sudden reduction of the population. 
(b) Temporal intensity profile of the laser pulse in MW cm$^{-2}$.}
\end{minipage}
\end{center}
\end{figure}

In Fig.~\ref{lasertrans} (a), the population evolution $P_{(16,15)}(t)$ of the state $(16,15)$ is plotted. 
At $t=7$ ns, a 1 ns-long resonant laser pulse of fluence $I\sim 6$ mJ cm$^{-2}$ and the temporal 
intensity profile shown in Fig.~\ref{lasertrans} (b) irradiates the atom and induces the transition 
$(16,15)\rightarrow (17,14)$. An abrupt 
$\varepsilon\sim 20\%$ reduction in $P_{(16,15)}(t)$ is seen, whereas the
$\pi^-$ excited to $(17,14)$ undergoes Auger decay at a rate $A_{(17,14)}=1.7\times 10^{11}$ s$^{-1}$. 
A corresponding spike appears in the rate of $\pi^-$ absorptions,
with an intensity of $\sim 10^{-3}$ normalized to the total number of $\pi^-$ which come to rest in 
the helium target at $t=0$. The simulation indicates that a fluence 
$I>30$ mJ cm$^{-2}$ can strongly saturate the transition and deplete most of the 
population in $(16,15)$. The high power requirement is due to the small transition amplitude, and
the large Auger rate of $(17,14)$, which causes rapid dephasing in the laser-induced transition.

Lasers of pulse length $\sim 10$ ns and lower fluence $I\sim 1$ mJ cm$^{-2}$ can saturate 
the favored transition $(17,16)\rightarrow (16,15)$ at a wavelength of 266.4 nm between 
two metastable states. This alone would not produce a spike in the rate of $\pi^-$ absorption events, 
since our spectroscopy method relies on inducing a transition to an Auger-dominated state. 
A second laser with saturating fluence tuned to $(16,15)$$\rightarrow$$(17,14)$ or 
$(17,16)$$\rightarrow$$(18,15)$ is needed to detect the change in the $\pi^-$ population 
induced by the first 266.4-nm laser, in a two-step resonance configuration.

\subsection{Resonance profile}
\label{profi}

\begin{figure}[tb]
\begin{center}
\includegraphics[height=10cm]{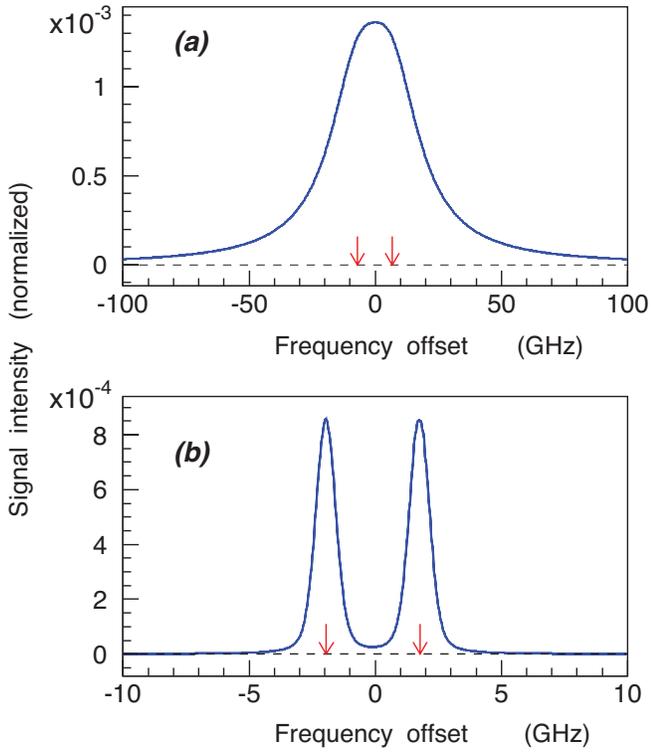}
\begin{minipage}[t]{8.5 cm}
\caption{\label{profile_resonance1} 
(Online color) Resonance profiles of the $\pi{\rm ^4He}^+$ transitions (a) $(n,\ell)=(16,15)\rightarrow (17,14)$
and (b) $(17,16)\rightarrow (16,15)$. Note the different scales for the laser frequency offset. 
Arrows indicate the positions of the dominant fine structure sublines.}
\end{minipage}
\end{center}
\end{figure}

In Fig.~\ref{profile_resonance1} (a), the resonance profile of the $\pi{\rm ^4He}^+$ transition
$(16,15)\rightarrow (17,14)$ excited with a laser fluence $I\sim 6$ mJ cm$^{-2}$ is simulated. The intensity 
of the $\pi^-$ absorption spike implied by Fig.~\ref{lasertrans} is plotted as a function of the laser 
frequency $\Delta\nu$ detuned from the spin-independent resonance frequncy $\nu_{\rm exp}$. 
The width of this profile is predominantly caused by the contribution $A_{(17,14)}/2\pi\sim 26$ GHz 
of the Auger decay rate of the resonance 
daughter state and the spacing ($\sim 13.7$ GHz) between the fine structure sublines, the positions 
of which are indicated by arrows. It should in principle be possible to determine the centroid of this profile with a precision of 
$<1$ GHz. This corresponds to a fractional precision of better than $\sim 1\times 10^{-6}$ on $\nu_{\rm exp}$.

Narrower lines are seen for the favored transition
$(17,16)\rightarrow (16,15)$ [Fig.~\ref{profile_resonance1} (b)] by using 266.4- and 383.8-nm
laser pulses in sequence. The frequency of the former laser is scanned over the resonance, whereas the latter 
is fixed to the transition $(16,15)$$\rightarrow$$(17,14)$. This profile was
simulated for a 10-ns-long laser pulse of $I\sim 0.1$ mJ cm$^{-2}$ for the 266.4-nm laser. The fine-structure
sublines can now be resolved as distinct peaks. The 1-GHz widths of the peaks arises
from the 7-ns-lifetime of $\pi{\rm He}^+$, power broadening effects, and 
the finite observation time. This implies that $\nu_{\rm exp}$ can in principle be determined to 
a precision of a few 10's MHz, which corresponds to a fractional precision of $\sim 10^{-8}$.

\subsection{Detection of $\pi^-$ absorption}
\label{absorption}

Past experiments on the longevity of $\pi^-$ in helium targets 
\cite{fetkovich1963,block1963,block1965,zaimidoroga1967,nakamura1992} used
bubble chambers, multiwire proportional chambers, and sodium iodide spectrometers 
to identify the varieties and trajectories of the particles that emerged from
the decay or nuclear absorption of $\pi^-$. This helped to isolate the signal electrons, $\mu^-$, or protons from 
background events. These techniques were optimized for low- to moderate count rates. 
A laser spectroscopy experiment of $\pi{\rm He}^+$, on the other hand, would need 
$\pi^-$ beam intensities and detection efficiencies which are $>2$ orders of magnitude higher, in order to resolve 
the resonance signal. We intend to achieve this by detecting the signal neutrons, protons, deuterons, and tritons 
using plastic scintillation counters \cite{soter2014} surrounding the experimental 
target, which do not have particle identification, or vertex reconstruction, capabilities.

The $\pi^-$ absorption into the ${\rm ^4He}$ nucleus predominantly leads to one of four final states \cite{schiff1961,eckstein1963,ziock1970,barrett1973,kubodera1975,cernigoi1981,daum1995,bistirlich1970,baer1977},
\begin{eqnarray}
\pi^- +{\rm ^4He}&\rightarrow& {\rm ^3H} +n + 118.5\ {\rm MeV}, \label{thirddecay}\\
                           &\rightarrow& {\rm ^2H}+ n + n + 112.2\ {\rm MeV}, \label{seconddecay}\\
                           &\rightarrow&\ \ p +n + n+ n+110.0\ {\rm MeV}, \label{firstdecay} \\
                           &\rightarrow& {\rm ^4H^{\ast}}+\gamma. \label{gammadecay}
\end{eqnarray}
Their measured branching ratios \cite{daum1995,bistirlich1970,baer1977} are respectively, 
$(17\pm 9)\%$, $(63\pm 26)\%$, $(21\pm 16)\%$, and $\sim 1.5\%$.

\begin{figure}[tb]
\begin{center}
\includegraphics[height=14cm]{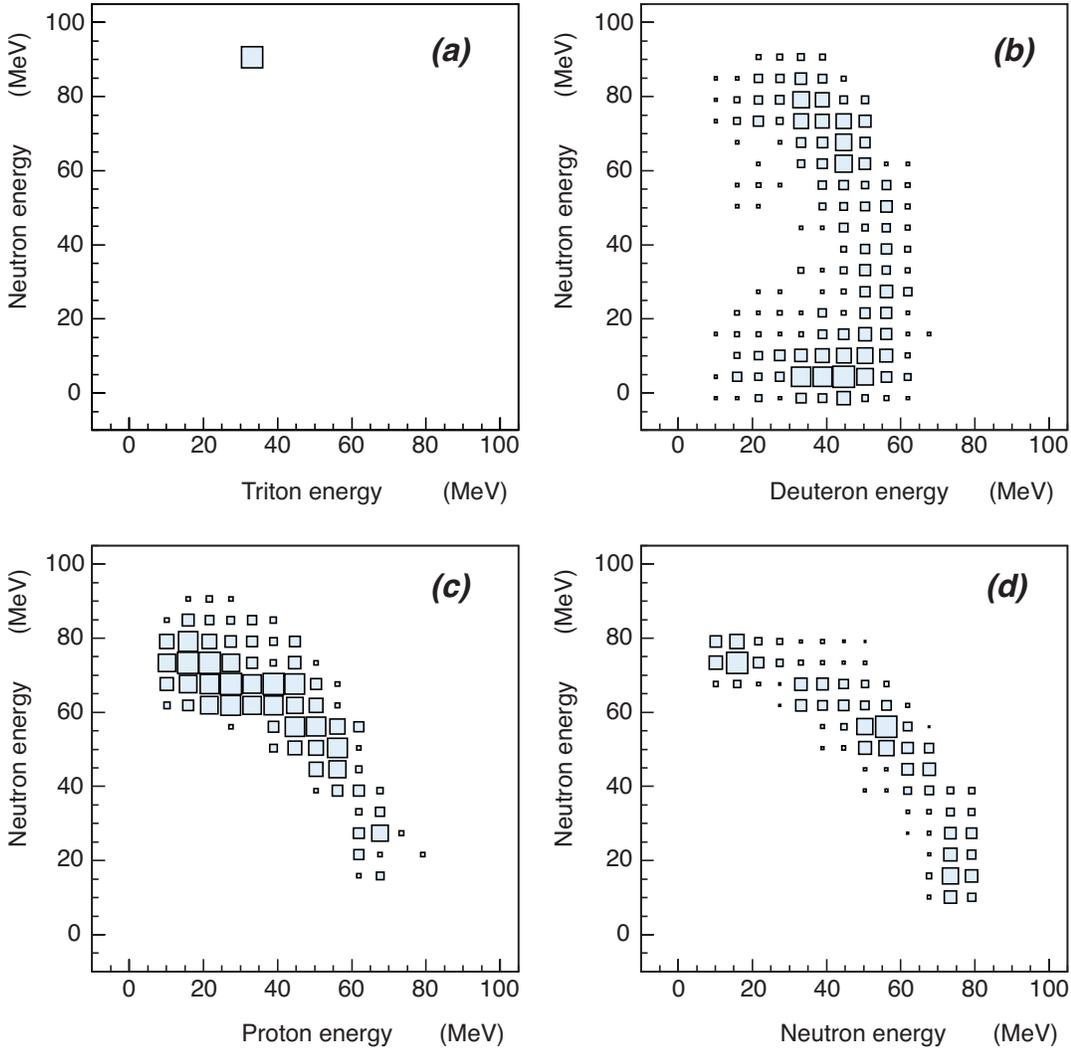}
\begin{minipage}[t]{8.5 cm}
\caption{\label{correlations} 
(Online color) Correlations of the kinetic energies of (a) triton-neutron, (b) deuteron-neutron, (c) proton-neutron, and
(d) neutron-neutron pairs which emerge from $\pi^-$ absorption into a $^4$He nucleus (see the text).
The primary energies of the particles before they slow down in the experimental apparatus are shown.
They are based on the experiments of Refs.~\cite{ziock1970,barrett1973,cernigoi1981,daum1995}, which 
are augmented by theoretical distributions 
\cite{eckstein1963,kubodera1975} for the regions where experimental data is not available. 
The $E\le 10$ MeV charged particles are ignored. Some of the details of these plots may be 
inaccurate, but they are sufficient to roughly simulate the intensity of the $\pi^-$ absorption signal 
in the $\pi{\rm ^4He}^+$ laser spectroscopy experiment.}
\end{minipage}
\end{center}
\end{figure}

Figs.~\ref{correlations} (a)--(c) show the correlations of the kinetic energies between 
triton-neutron, deuteron-neutron, and proton-neutron pairs that emerge in the non-radiative channels of 
Eqs.~\ref{thirddecay}, \ref{seconddecay}, and \ref{firstdecay}, respectively. 
Fig.~\ref{correlations} (d) shows the neutron-neutron pairs which originate from the channels 
of Eqs.~\ref{seconddecay} and \ref{firstdecay} added together. All these particles tend to 
emerge collinearly \cite{daum1995}. The plots are based on the experiments of 
Refs.~\cite{ziock1970,barrett1973,cernigoi1981,daum1995},
which are augmented by theoretical distributions \cite{eckstein1963,kubodera1975} for 
regions where experimental data are not available. The $E\le 10$ MeV charged particles are ignored.
Some details of these plots may be inaccurate, but they are sufficient to roughly simulate 
the $\pi{\rm He}^+$ signal. 

In the two-body channel of Eq.~\ref{thirddecay}, a monoenergetic triton and neutron of 
energies $E_t=30.6$ MeV and $E_n=90$ MeV [Fig.~\ref{correlations} (a)] emerge back to back.
The energy distribution [Fig.~\ref{correlations} (b)] of the three-body channel of 
Eq.~\ref{seconddecay} has been interpreted by several groups in the following way 
\cite{kubodera1975,daum1995}. 
Due to final-state interaction effects, the spectrum contains a strong peak corresponding to a $E_d\sim 56$ MeV 
deuteron, which is accompanied by two collinear $E_n\sim 28$-MeV neutrons emitted in the opposite direction. 
The two maxima of Fig.~\ref{correlations} (b) at positions of
$(E_d,E_n)\sim (40 {\rm \ MeV}, 70 {\rm \ MeV})$ and $(40 {\rm \ MeV}, 0 {\rm \ MeV})$ correspond 
to the so-called 
``quasi-free" absorption of $\pi^-$. This leads to the deuteron and a neutron emitted nearly back-to-back, 
while the remaining neutron plays the role of a spectator of low energy.
The peak at $E_n=56$ MeV in the neutron spectrum of Fig.~\ref{correlations} (d) 
represents the quasi-free absorption of $\pi^-$ into a proton-neutron pair within the nucleus;
this results in the emission of two back-to-back neutrons and a spectator deuteron.
The minimum near the center of the truncated ellipse of Fig.~\ref{correlations} (b) at 
$(E_d,E_n)\sim (30 {\rm \ MeV},40 {\rm \ MeV})$ is the unlikely case where the three 
particles emerge in a non-collinear way, with momenta which are uniformly distributed in phase space.

The energy distribution of the protons in the four-body channel of Eq.~\ref{firstdecay} extends
up to $E_p\sim 70$ MeV and is skewed towards lower energies [Fig.~\ref{correlations} (c)]. 
The quasi-free absorption of the $\pi^-$ into a proton-proton pair within the nucleus 
can lead to the back-to-back emission of a proton and a neutron of $E_n\sim E_p\sim 56$ MeV. The 
measured branching ratio of this process is small $(3\pm 1)\%$ \cite{daum1995}. 

Equation~\ref{gammadecay} describes those $\pi^-$ that occupy the low-$n_i$ 
ionic states of $\pi{\rm He}^{2+}$, that subsequently undergo radiative capture 
\cite{bistirlich1970,baer1977} into $^4{\rm He}$ nuclei. 
The energy spectrum of the emitted $\gamma$-rays has a maximum at $E_{\gamma}\sim 112$ MeV. 
Its shape implies the excitations of several resonance states of the ${\rm ^4H}$ nucleus, at 
energies of $E=3.4$--7.4 MeV relative to the ${\rm ^3H} +n$ decay threshold. Low-energy 
neutrons are subsequently emitted.

The final states for $\pi^-$ absorption into the ${\rm ^3He}$ nucleus
\cite{truol1974,backenstoss1986,corriveau1987,gotta1995} are predominantly, 
\begin{eqnarray}
\pi^- +{\rm ^3He} &\rightarrow& {\rm ^2H}+ n, \label{pi3_1} \\
                                 &\rightarrow& \  \ p +n + n, \label{pi3_2} \\
                                 &\rightarrow&  {\rm ^3H} +\gamma, \label{gamma1}\\
                                 &\rightarrow&  {\rm ^2H} +n+\gamma, \label{gamma2}\\
                                 &\rightarrow&  \ \ p+n+n+\gamma, \label{gamma3} \\
                                 &\rightarrow& {\rm ^3H}+\pi^0 \label{pi_0}.
\end{eqnarray}
The experimental branching ratios are respectively $(16\pm 2)\%$, $(58\pm 5)\%$,
$(5.7\pm 0.2)\%$, $(3.6\pm 1.2)\%$,
$(3.6\pm 1.3)\%$, and $(15\pm 0.6)\%$ \cite{corriveau1987,gotta1995}.
The nonradiative nuclear absorptions of Eqs.~\ref{pi3_1} and \ref{pi3_2} are expected to produce 
the strongest signals in the $\pi{\rm ^3He}^+$ experiment. The former results in the back-to-back 
emission of a monoenergetic deuteron 
and a neutron of $E_d\sim 45$ MeV and $E_n\sim 90$ MeV \cite{gotta1995}. The protons and neutrons 
in the three-body channel of Eq.~\ref{pi3_2} are
distributed up to $>90$ MeV and tend to emerge collinearly. Unlike the ${\rm ^4He}$ case, there are significant
contributions from radiative nuclear capture (Eqs.~\ref{gamma1}--\ref{gamma3}) and charge-exchange 
(Eq.~\ref{pi_0}) processes between the $\pi^-$ and ${\rm ^3He}$ nucleus.

When $\pi^-$ are allowed to come to rest in thick ${\rm ^4He}$ targets, some of the secondary protons, 
deuterons, and tritons slow down and stop in the helium. This reduces the efficiency $D$ of the 
scintillation counters detecting the $\pi^-$ absorption. The experiment of 
Ref.~\cite{nakamura1992}, for example, counted the $E_p>50$-MeV protons which emerged from a
liquid helium target of diameter $d=230$ mm. This provided a clean signal for delayed $\pi^-$ events with a
high degree of background elimination, but we estimate the corresponding $D$-value to be 
$<1\%$. This efficiency can be increased by  (i) using a small ($d<40$ mm) helium target which
allows protons, deuterons, and some tritons of $E_p\ge 20$ MeV, $E_d\ge 30$ MeV, and $E_t=30.6$ MeV 
to emerge from it and (ii) detecting the neutrons that are produced at
an order-of-magnitude higher rate than the protons. A plastic scintillation counter of thickness 
$t_r\sim 40$ mm can detect $E_n=10$-, 40-, and 80-MeV neutrons with efficiencies of 
$\sim 10\%$, $\sim 5\%$, and $\sim 3\%$, respectively \cite{abdel1982}. 

\subsection{Coincidences and event rates}
\label{eventrates}

The arrival of the laser pulses at the experimental target must coincide with
the formation of $\pi{\rm He}^+$, if any transition is to be induced. This is difficult to achieve since
the $\pi{\rm He}^+$ decays within tens of nanoseconds, whereas the arrival time of 
individual $\pi^-$ produced in synchrotron facilities cannot easily be predicted with 
nanosecond-scale precision. Past experiments on
 $\overline{p}{\rm He}^+$ \cite{Mor94,mhori2004} or muonic hydrogen atoms \cite{pohl2010}
first detected the formation of the exotic atoms and
then triggered some pulsed lasers to irradiate them at a time $t=1$--10 $\mu{\rm s}$ 
after formation. In the $\pi{\rm He}^+$ case, the detected atoms would decay 
well before the laser pulses could reach them.
Another method for $\overline{p}{\rm He}^+$ spectroscopy \cite{mhori2004,mhori2011}
involved ejecting a 200-ns-long pulsed beam containing $>10^7$
antiprotons from a synchrotron and triggering the laser so that a single laser pulse 
arrived at the target with a $t=1$--10 $\mu{\rm s}$ delay, relative to the antiprotons. 
A bright 200-ns-long flash caused by the antiprotons that promptly annihilated 
in the helium was detected, followed by a much longer but less intense tail from the delayed 
annihilations of the metastable $\overline{p}{\rm He}^+$.
The laser resonance signal was superimposed on this tail. This method is also difficult to apply to 
$\pi{\rm He}^+$, which would be destroyed well within the initial flash of nuclear absorptions 
that occurs during the arrival of a 200-ns-long $\pi^-$ beam.

We instead propose synchronization of the laser pulses to the $\pi^-$ beam of the PSI cyclotron 
\cite{seidel2010}. Here radiofrequency cavities excited at $f_c=50.63$ MHz 
arrange the protons into 0.3-ns-long bunches at intervals of $f_c^{-1}=19.75$ ns, and
accelerate them to a kinetic energy of $E=590$ MeV. The protons are then allowed to collide with a 40-mm-thick 
carbon target, thereby producing the $\pi^-$. 
A magnetic beamline \cite{meg2013} collects the $\pi^-$ of a certain 
momentum, and transports them over $\sim 15$ m to the position of the experimental 
helium target. The $\pi^-$ thus arrive as a train of bunches 
that are synchronized to $f_c$. This signal triggers the laser, so that the laser pulse
arrives $t=7$ ns after a $\pi^-$ bunch. Most of the laser pulses will irradiate an ``empty" target, 
since the $\sim 98\%$ majority of the $\pi^-$ are
absorbed promptly without forming metastable atoms \cite{nakamura1992}. A high $\pi^-$ beam intensity of 
$N_{\pi}>f_c=5\times 10^7$ s$^{-1}$ is needed to ensure a sufficient ($>0.1\%$) probability of coincidence 
between the laser pulses and $\pi{\rm He}^+$ populating the resonance parent state. 

The rate of detecting $\pi^-$ absorption events which signal the laser transition can then be estimated as
\begin{equation}
\Gamma_{\rm evt}\sim\varepsilon DN_{\pi}S_\pi P_{(n,\ell)}f_{\rm las}f_c^{-1}.
\label{ddd}
\end{equation}
According to the simulation of Sec.~\ref{stopping}, $S_\pi>50\%$ of the incoming $\pi^-$ 
come to rest in a 20-mm-diam volume of the helium target. This volume is then irradiated by a
resonant laser beam of repetition rate $f_{\rm las}\sim$ 0.1--1 kHz.
Only the data from $\pi^-$ bunches that coincide with the laser are acquired at a rate 
$f_{\rm las}$,  whereas the remaining bunches of rate $f_c-f_{\rm las}$ are ignored. 
Among these $\pi^-$, the fraction that occupies the resonance parent state at the moment of laser irradiation
is assumed in Sect.~\ref{cascsect} and Fig.~\ref{lasertrans} to be $P_{(16,15)}(t=7{\ \rm ns})\sim 0.007$. 
The efficiency of the laser depopulating the parent state is estimated in Sect.~\ref{tranprob} to be $\varepsilon \sim 20\%$. The resulting $\pi^-$ absorptions are detected by the scintillation counters with a typical efficiency of $D\sim 20\%$. A $\pi^-$ beam intensity of $N_{\pi}\sim 10^8$ s$^{-1}$ would therefore imply $\Gamma_{\rm evt}\sim 100-1000$ h$^{-1}$.

\section{Monte-Carlo simulations}
\label{simulations}

\subsection{Stopping of $\pi^-$ in helium}
\label{stopping}

\begin{figure}[tb]
\begin{center}
\includegraphics[height=12cm]{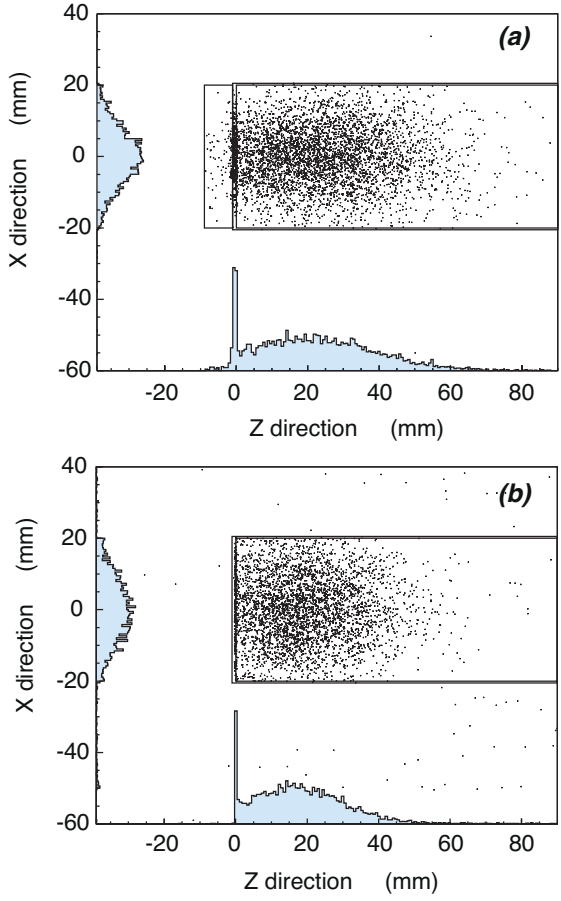}
\begin{minipage}[t]{8.5 cm}
\caption{\label{fig:stopping} 
(Online color) (a) Simulated spatial distribution of $\pi^-$ with momentum 80 MeV/c and momentum spread $\sim 8\%$ coming to rest in a liquid-helium target. A two-dimensional projection, and $x$
and $z$ projections of the $\pi^-$ distributions are shown 
superimposed. The $\pi^-$ arrive from the left side of the diagram and traverse an 8-mm-thick aluminum 
degrader foil before entering the 40-mm-diameter target at position $z=0$. The outlines of the foil and target chamber 
are indicated by solid lines. (b) Spatial stopping distribution of 40-MeV/c $\pi^-$ beam, 
coming to rest in a helium gas target of pressure 
$p\sim 1.7\times 10^5$ Pa and temperature $T\sim 6$ K (b).}
\end{minipage}
\end{center}
\end{figure}

The $\pi^-$ must come to rest in the helium target with a high efficiency, 
to maximize the produce yield of $\pi{\rm He}^+$.
The $\pi$E5 beamline of PSI \cite{meg2013}
provides a $\pi^-$ beam of intensity $N_{\pi}\sim 10^8$ s$^{-1}$, momentum 
$\sim 80$ MeV/c, momentum spread $\sim 8\%$, and diameter $d\sim 20$ mm.
Figure~\ref{fig:stopping} (a) shows the simulated spatial distribution of the $\pi^-$
coming to rest in a $40$-mm-diameter, 100-mm-long cylindrical chamber filled with 
liquid helium of atomic density $\rho\sim 2\times 10^{22}$ cm$^{-3}$. The slowing down of the $\pi^-$
is simulated using
the multiple scattering effect according to the Moli\`{e}re distribution, and the energy straggling effect 
according to the Vavilov distribution. The $\pi^-$ traverse an aluminum degrader plate 
before entering the helium target; 
by carefully adjusting the degrader thickness to $t_r\sim 8$ mm, some $S_{\pi}>50\%$ of the $\pi^-$ 
are stopped within a 20-mm-diameter, 50-mm-long cylindrical region along the beam axis. 
The small size of this stopping distribution ensures a high timing resolution and 
detection efficiency for $\pi^-$ absorption. It also allows the $\pi{\rm He}^+$ produced in this volume
to be readily irradiated with high-intensity laser light, fired into the target in an anticollinear 
direction with the $\pi^-$ beam, as in the experimental setups of Refs.~\cite{mhori2001,mhori2004}.

Beams of lower momentum $\pi^-$ can be efficiently stopped in a helium-gas target, 
where collisional quenching, broadening, and shifting of the $\pi{\rm He}^+$
resonance lines may be reduced.
Fig.~\ref{fig:stopping} (b) shows the spatial distribution of $\pi^-$ with a momentum of 40 MeV/c 
which come to rest in a gas target of pressure $p\sim 1.7\times 10^5$ Pa and temperature 
$T\sim 6$ K. This corresponds to a density $\rho\sim 2.6\times 10^{21}$ cm$^{-3}$.
The intensity of such a beam is limited to $N_{\pi}\le 10^6$ s$^{-1}$,
due to the lower $\pi^-$ production yield, and the fact that most of the slow $\pi^-$ decay in-flight before they can 
reach the helium target. This implies a signal rate $\Gamma_{\rm evt}\sim1-10$ h$^{-1}$ according to Eq.~\ref{ddd}. 

\subsection{Backgrounds}
\label{possback}

The experiment has several background sources:

 {\it Prompt absorption,} 
the 98$\%$ majority of the $\pi^-$ that come to rest in the helium target undergo
nuclear absorption within picoseconds (Sect.~\ref{absorption}). 
The flux of neutrons and charged particles produced by 
this is $>10^3$ times larger than the laser spectroscopic signal of $\pi{\rm He}^+$. 
The signal can be isolated from this background, by adjusting the laser pulse to
arrive $t\sim 7$ ns after the $\pi^-$ (see Fig.~\ref{lasertrans}).

{\it Cascade of $\pi{\rm He}^+$,}  only a small fraction of the metastable $\pi{\rm He}^+$ will
undergo laser excitation. The remaining atoms will decay spontaneously, 
and produce a continuous background in the measured time spectrum.

{\it Electron contamination,} the $\pi^-$ beam is contaminated with electrons of intensity 
$N_e\sim 10^9$ s$^{-1}$. These electrons can scatter off the experimental target and produce 
background in the surrounding scintillation counters. The contamination can be reduced to 
$N_e\sim 10^7$ s$^{-1}$ by using a $E\times B$ separator positioned upstream of the target.

{\it Muon contamination,} the beam also contains contaminant $\mu^-$ with a ratio 
$N_{\mu}/N_{\pi}\sim 1$, depending on the momentum and flight path of the $\pi^-$. 
Some of the $\mu^-$ come to rest in the helium target and 
produce muonic helium atoms. Most of these undergo 
$\mu^-\rightarrow e^-+\overline{\nu}_e+\nu_{\mu}$ decay into an electron, an electron-based
antineutrino, and a muon-based neutrino with a partial decay rate
of $\gamma_{\mu}\sim 4.5\times 10^5$ s$^{-1}$. The
electrons of mean energy $E\sim 40$ MeV can be isolated to some extent from the 
signal neutrons, protons, deuterons, and tritons by their smaller energy loss in the 
scintillation counters \cite{pdg}.
The competing process of $\mu^-$ capture into the nucleus via weak interaction,
\begin{equation}
\mu^-+{\rm ^4He}\rightarrow {\rm ^3H}+n+\nu_{\mu},
\end{equation}
has a small capture rate $\gamma_{\rm cap}=300$--400 s$^{-1}$ \cite{measday2001}, 
so that the emerging neutrons do not constitute a significant background. 
The $\mu^-$ that stop in the metallic walls of the target chamber are
captured into heavier nuclei at a higher rate
$\gamma_{\rm cap}>7\times 10^4$ s$^{-1}$ \cite{measday2001}. 
This results in the emission of 0.1--0.2 neutron of energy
$E_n\ge 10$ MeV per stopped $\mu^-$ \cite{measday2001}. Since most of these neutrons are of
$E_n<20$ MeV, this background can be reduced by rejecting those events
with a small energy deposition in the scintillation counters.

\subsection{Signal-to-background ratios}

The signal-to-background ratio of the laser resonance depends on 
the design of the experimental apparatus, the characteristics of the $\pi^-$ and laser beams,
the populations in the $\pi{\rm He}^+$ states, and the data analysis methods. 
We carried out Monte-Carlo simulations using the GEANT4 toolkit \cite{geant1}, to roughly 
estimate the signal-to-background ratio for an apparatus that is conceptually similar to 
those of Refs.~\cite{mhori2001,mhori2004}. The $\pi^-$ beam of momentum 
80 MeV/c came to rest in a liquid-helium target (Sec.~\ref{stopping}) at intervals 
of $f_c^{-1}=19.75$ ns. The beam contained contaminant electrons and $\mu^-$ with ratios 
of $N_e/N_{\pi}=0.2$ and $N_{\mu}/N_{\pi}=1$.
Some $1.6\%$ and $0.7\%$ of the $\pi^-$ (Sec.~\ref{cascsect}) were respectively captured into the 
$\pi{\rm He}^+$ states $(n,\ell)=(16,15)$ and $(17,16)$. These
atoms decayed with a mean lifetime of $\tau_c\sim 7$ ns. 
The laser transition $(16,15)$$\rightarrow$$(17,14)$ induced at
$t=7$ ns depopulated $\varepsilon\sim 20\%$ of the $\pi^-$ that occupied $(16,15)$
(Sect.~\ref{tranprob}). Neutrons, protons, deuterons, and tritons emerged from the resulting
$\pi^-$ absorption with the branching ratios and energy distributions described in
Sect.~\ref{absorption}. For simplification, we simulated only the two highest-energy
particles in each event of Eqs.~\ref{seconddecay} and \ref{firstdecay}. Due to the dominant 
nature of the quasi-free absorption \cite{daum1995}, we assume that the remaining particles 
tend to have low energies and would not strongly contribute to the $\pi{\rm He}^+$ signal. 
Some of these signal particles traversed the 1.4-mm-thick lateral walls of the target chamber
(Fig.~\ref{fig:stopping}) made of aluminium. They were detected by an array of 40-mm-thick plastic 
scintillation counters (Sect.~\ref{absorption}), which covered a solid angle of $\sim 2\pi$ steradians 
seen from the target. The trajectories of the particles were followed for $\sim 100$ ns. 

\begin{figure}[tb]
\begin{center}
\includegraphics[height=11.5cm]{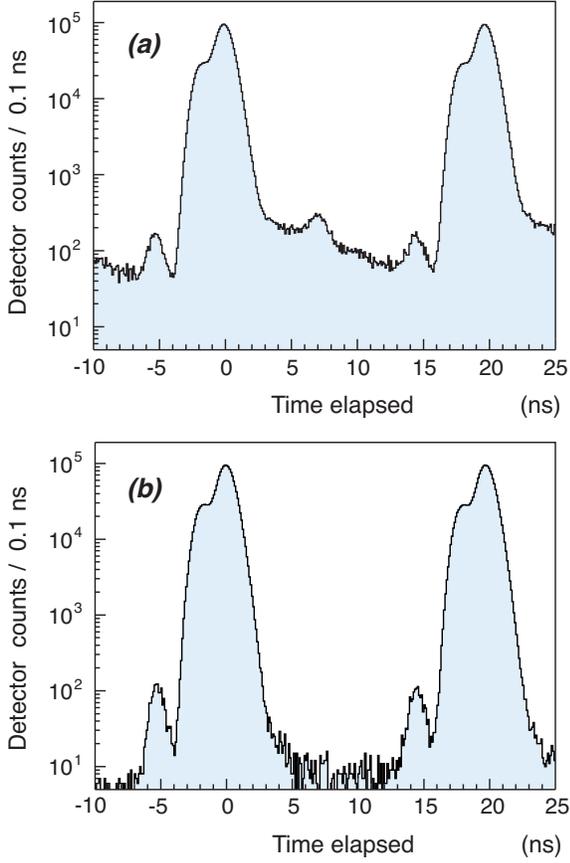}
\begin{minipage}[t]{8.5 cm}
\caption{\label{fig:adats_pion} 
(Online color) (a) Simulated time spectrum of metastable $\pi{\rm ^4He}^+$. 
The time elapsed from $\pi^-$ arrival at a liquid helium target, until a particle hit was 
registered in an array of plastic scintillation counters surrounding the target is shown (see the text). 
The $\pi^-$ arrive at intervals of $f_c^{-1}=19.75$ ns, synchronized to the 
acceleration frequency $f_c$ of the PSI cyclotron. The laser resonance transition 
$(n,\ell)=(16,15)\rightarrow (17,14)$ is induced at $t=7$ ns. The resulting peak 
in the rate of particle hits is visible over the background arising from the spontaneous 
decay of $\pi{\rm ^4He}^+$. (b) Time spectrum obtained by assuming that 
all the $\pi^-$ are promptly absorbed in the helium target, without forming metastable 
$\pi{\rm ^4He}^+$.}
\end{minipage}
\end{center}
\end{figure}

Fig.~\ref{fig:adats_pion} (a) shows the spectrum of time elapsed from $\pi^-$ arrival at the experimental target,
until a hit was registered in the scintillation counters. It represents $10^7$ $\pi^-$ arrivals coinciding with 
the laser pulse. Some of the background electrons and neutrons were rejected by accepting only 
those events that deposited more than $E_{\rm cut}=10$ MeV of
energy in the counters. No other particle identification or vertex reconstruction was carried out. The 
peaks at $t=0$ and $19.75$ ns correspond to arrivals of $\pi^-$ which undergo
prompt nuclear absorption. The secondary peaks at $t=-2$ and  $18$ ns represent
$\pi^-$ that stop in the upstream walls of the target chamber. Those at $t=-5.5$ and $14.5$ ns are
caused by the electrons that scatter off the target and strike the scintillation counters. All these
peaks are broadened due to the finite timing resolution of the scintillation counters which
was assumed to be $\sim 1$ ns and the time-of-flight of the particles in the experimental 
apparatus.

The spontaneous deexcitations of metastable $\pi{\rm He}^+$ give rise to 
the continuous spectrum which decays with a lifetime of $\tau_c=7$ ns. The peak at $t=7$ ns with a signal-to-background ratio of $\sim 2$ corresponds to the laser resonance. The intensity of this signal peak is linearly 
dependent on the population $P_{(16,15)}(t)$ of the parent state, and the 
$\varepsilon$-value of the laser (Sect.~\ref{tranprob}).
The spectrum represents $\sim 30$ h of data taking at the 
experimental conditions of Sect.~\ref{eventrates}; it may therefore take a few months to repeat the measurements 
at 20--30 settings of the laser frequency around the $\pi{\rm He}^+$ resonance, and obtain the spectral 
profile of Fig.~\ref{profile_resonance1} needed to determine $\nu_{\rm exp}$.

Fig.~\ref{fig:adats_pion} (b) shows the time spectrum obtained by assuming that all the $\pi^-$ are 
promptly absorbed in the helium without forming metastable $\pi{\rm He}^+$. This spectrum represents 
the sum of all the background processes of Sect.~\ref{possback}. Its intensity critically 
depends on the design of the target and scintillation counters, and the $E_{\rm cut}$-value used 
to reject the minimum-ionizing electrons that emerge from, e.g., 
$\mu^-\rightarrow e^-+\overline{\nu}_e+\nu_{\mu}$ decay. These must be carefully
optimized in the actual experiment.

\section{Discussions and Conclusions}
\label{conclude}

We have carried out three-body QED calculations on the spin-independent energies, 
fine structure, $E1$ transition frequencies, and Auger and radiative rates of metastable 
$\pi{\rm ^3He}^+$ and 
$\pi{\rm ^4He}^+$ atoms in the region $n=15$--19. We described a method for 
laser spectroscopy, in which pulsed lasers induce transitions from the metastable 
states, to states with picosecond-scale lifetimes against Auger decay.
The Rydberg $\pi{\rm He}^{2+}$ ion that remains after Auger emission
is rapidly destroyed in collisions with helium atoms. Here the ionic states
undergo Stark mixing with $s$, $p$, and $d$ states, which leads to $\pi^-$ absorption 
into the helium nucleus. This results in a spike in the rates of neutrons, protons, deuterons, and tritons 
which emerge with kinetic energies up to 30--90 MeV. This signals the resonance condition between the laser and 
$\pi{\rm He}^+$. The feasibility of this experiment depends on the population of $\pi{\rm He}^+$ that 
occupies the resonance parent state. Based on the experimental work on $\overline{p}{\rm He}^+$,
we assume that most of the metastable population lies in the states $(n,\ell)=(16,15)$ and
$(17,16)$. We found that a cascade model which includes assumptions on collisional 
deexcitation rates, can reproduce the features of past experimental data \cite{nakamura1992} on the lifetime of 
$\pi^-$ in liquid helium targets. Several candidate transitions, e.g., $(16,15)\rightarrow (17,14)$, 
appear to be amenable for detection. Although the decay and nuclear absorption of $\pi^-$ and $\mu^-$ 
give rise to several experimental backgrounds, Monte-Carlo simulations suggest that a 
spectroscopic signal can be resolved, if appropriate particle detection techniques were used. 
The PiHe collaboration now attempts to carry out this measurement using the PSI ring cyclotron facility. 
The experiment involves allowing a $\pi^-$ beam of intensity $>10^8$ s$^{-1}$ to come to rest in a helium target.
The time structure of the beam will be used to synchronize the formation of $\pi{\rm He}^+$ 
in the target, with the arrival of resonant laser beams of fluence 0.1--10 mJ$\cdot$cm$^{-2}$. 
This may allow the transition frequencies of $\pi{\rm He}^+$ to be determined with a fractional 
precision of $10^{-8}-10^{-6}$ provided that the systematic uncertainties can be controlled.

Several effects can prevent the detection of the laser resonance signal, beyond those evaluated in this paper.
Experiments on $\overline{p}{\rm He}^+$ have shown that some states become short-lived to nanosecond- 
and picosecond-scale lifetimes when the density of the helium target is increased 
\cite{mhori1998,mhori1998e,mhori2001,mhori2004}. Collisions between $\overline{p}{\rm He}^+$ and 
normal helium atoms broadened and weakened some laser resonances, so that they could no longer be detected . 

The mass $M_{\pi^-}$ of the $\pi^-$ can be determined by comparing the measured frequencies $\nu_{\rm exp}$ of
$\pi{\rm He}^+$ with the results of the three-body QED calculations $\nu_{\rm th}$ described in this paper. 
The precision on $M_{\pi^-}$ will be ultimately limited by the finite lifetime $\tau_{\pi}\sim 26$ ns of $\pi^-$ 
according to the expression, $\left(2\pi \tau_{\pi}\nu_{\rm exp}\theta\right)^{-1}$, since this determines the 
relative natural width of the resonance compared to its energy.  Here $\theta=1$--9 denotes the
sensitivity of $\nu_{\rm th}$ on $M_{\pi^-}$. When $M_{\pi^-}$ was increased by 1 part per billion (ppb)
in our three-body calculations, the $\nu_{\rm th}$ values for the $\pi{\rm ^4He}^+$ transitions
$(17,16)\rightarrow (16,15)$ and $(16,15)\rightarrow (17,14)$,
and the $\pi{\rm ^3He}^+$ transitions
$(17,16)\rightarrow (16,15)$ and $(16,15)\rightarrow (17,14)$
were found to shift by $1.9$, $-1.1$, $1.9$, and $-9$ ppb, respectively \cite{korobov2006,korobov2008}.
The last of these transitions appears to be particularly sensitive to $M_{\pi^-}$.
This implies that a fractional precision on $M_{\pi^-}$ of $<10^{-8}$
can in principle be achieved, as in the case of $\overline{p}{\rm He}^+$ \cite{mhori2011,mhori2006}. In practice, 
systematic effects such as the shift and broadening of the resonance line due to atomic collisions 
\cite{mhori2001,bakalov,adamczak2013,kobayashi2013} 
in the experimental target, AC Stark shifts \cite{mhori2010}, frequency chirp in the laser beam \cite{mhori2009},
and the statistical uncertainties due to the small number of detected events can prevent the experiment from achieving this precision.  
The lifetimes of some $\pi{\rm He}^+$ states may be shorter than $\tau_{\pi}$, due to collisional effects (Sects.~\ref{cascsect} and \ref{profi}).

Past values of $M_{\pi^{-}}$ have been determined \cite{pdg} in two ways, 
i): the X-ray transition energies of $\pi^-$ atoms were measured, and the results compared with relativistic bound-state calculations \cite{marushenko1976,carter1976,lu1980,jeckelmann1986,jeckelmann,lenz}, or ii): the recoil momentum 
$p_{\mu^+}=29.79200(11)$ MeV/c \cite{abela,daum,assamagan1994,assamagan1996} of the positively-charged muon $\mu^+$
which emerges from a stationary, positively-charged pion undergoing $\pi^+\rightarrow\mu^++\nu_{\mu}$ decay was precisely
measured. The relativistic kinematical formula,
\begin{equation}
p_{\mu^+}^2+m_{\mu^+}^2=\left(M_{\pi^{+}}^2+m_{\mu^+}^2-m^2_{\nu_\mu}\right)^2/4M_{\pi^{+}},
\label{pmu}
\end{equation}
was employed to determine the mass $M_{\pi^{+}}$ of the $\pi^+$, which according to CPT symmetry is equivalent to
$M_{\pi^{-}}$. The mass $m_{\mu^+}$ of the $\mu^+$ was determined to a fractional precision of 
$4\times 10^{-8}$ from the results of microwave spectroscopy of the ground-state hyperfine structure of 
muonium atoms \cite{liu1999}; the $\nu_{\mu}$ was assumed to be very light 
($m_{\nu_\mu}\sim 0$). The X-ray spectroscopy measurements are conventionally \cite{pdg} treated as 
direct laboratory determinations of $M_{\pi^-}$, as they involve no such assumption on $m_{\nu_\mu}$.

The $M_{\pi^-}$ value with the highest fractional precision of $\sim 3\times 10^{-6}$ was 
obtained from measurements of the 4f--3d transition energy of $\pi{\rm ^{24}Mg}$ atoms \cite{jeckelmann1986,jeckelmann,pdg}.
The results of these experiments have shown a bifurcation, i.e., 
two groups of results near 139.570 and 139.568 MeV/$c^2$ \cite{pdg} which arise 
from different assumptions on the electrons occupying the K-shell of the atom during X-ray emission. 
Ref.~\cite{pdg} notes that although the two solutions are equally probable, they chose the 
higher $M_{\pi^-}$ value which is consistent with a positive mass-squared value for $\nu_{\mu}$, i.e., 
$m^2_{\nu_\mu}>0$, according to Eq.~\ref{pmu}. 
The lower $M_{\pi^-}$ value would result in a so-called tachionic solution 
$m^2_{\nu_\mu}<0$. A separate spectroscopy measurement on the 5g--4f transition in
$\pi{\rm N}$ atoms \cite{lenz} which did not suffer from the ambiguity on the electron shell occupancy, 
determined $M_{\pi^-}$ with a factor $\sim 1.5$ lower precision;  the result was consistent with the 
higher $M_{\pi^-}$ value. 

The present limit \cite{pdg} on the $\nu_{\mu}$ mass obtained from direct laboratory measurements 
is $m_{\nu_\mu}<190$ keV/$c^2$ with a confidence level of 90$\%$.
This result was obtained by combining the $\pi^-$ and $\mu^+$ masses determined
by the above spectroscopy experiments on exotic atoms,
the recoil momentum $p_{\mu^+}$ \cite{abela,daum,assamagan1994,assamagan1996}, and Eq.~\ref{pmu}. 
Although cosmological observations and neutrino oscillation experiments 
have established a much better limit of $\Sigma m_{\nu}<0.2-0.4$ eV/$c^2$ for the sum of the masses of stable
Dirac neutrinos \cite{pdg}, this still represents the best limit obtained from direct laboratory measurements.
By improving the experimental precision on $M_{\pi^-}$, the direct limit on $m_{\nu_\mu}$ can be 
improved by a factor $\sim 2$. Further progress would require a higher-precision measurement of 
$p_{\mu^+}$ using, e.g., the muon $g$-2 storage ring of Fermilab \cite{carey2000}.

\begin{acknowledgments}
We are indebted to D.~Barna and the PSI particle physics and beamline groups for 
discussions. This work was supported by the European Research Council (ERC-StG) and the 
Heisenberg-Landau program of the BLTP JINR.
\end{acknowledgments}

\appendix

\section{Relativistic and radiative corrections}

We briefly describe the relativistic and radiative corrections in our calculations. 
Since we aim for $\sim 7$ significant digits of precision on the $\pi{\rm He}^+$ energy, 
recoil effects may be neglected. Atomic units are used.

The leading $R_\infty\alpha^2$-order relativistic contribution of the bound electron 
in the field of two massive particles can be expressed as \cite{BS},
\begin{equation}
E_{rc} = \alpha^2
   \left\langle
      -\frac{\mathbf{p}_e^4}{8m_e^3}
      +\frac{1+2a_e}{8m_e^2}
         \left[
            Z_{\rm He}4\pi\delta(\mathbf{r}_{\rm He}^{})
            +Z_{\pi}4\pi\delta(\mathbf{r}_{\pi})
         \right]
   \right\rangle.
\end{equation}
Here $a_e=1.1596522\times 10^{-3}$ denotes the anomalous magnetic moment of the electron.
The expectation values of the operators ${\bf p}^4_e$, $\delta({\bf r}_{_{\rm He}})$, and $\delta({\bf r}_{\pi})$
for $\pi{\rm ^4He}^+$ and $\pi{\rm ^3He}^+$ states are shown in Tables \ref{nonrela} and \ref{nonrela2}.

The next largest contribution is the one-loop self-energy of order $R_\infty\alpha^3$,
\begin{eqnarray}
E_{se}^{(3)} &=& \alpha^3\frac{4}{3}
   \left[\ln{\frac{1}{\alpha^2}}-\beta(L,v)+\frac{5}{6}-\frac{3}{8}\right] \nonumber \\
   &\times & \left\langle
      Z_{\rm He}\delta(\mathbf{r}_{\rm He}^{})+
      Z_{\pi}\delta(\mathbf{r}_{\pi})
   \right\rangle,
\end{eqnarray}
where $\beta(L,v)$ denotes the Bethe logarithm of the mean excitation energy due to emission and re-absorbtion of a virtual photon \cite{BS,PRA13RBL,korobov2014}. We used the adiabatic effective potential $\beta_{nr}(R)$ (see Fig.~4 in Ref.~\cite{PRA13RBL}) for the two-center problem with Coulomb charges $Z_1=2$, $Z_2=-1$, which was averaged over the vibrational wave function $\chi_{L,v}^{}(R)$ of a particular state. We also included the one-loop vacuum polarization using the expression,
\begin{equation}
E_{vp}^{(3)} =\frac{4\alpha^3}{3}
   \left[-\frac{1}{5}\right] \Bigl\langle
      Z_{\rm He}\delta(\mathbf{r}_{\rm He}^{})+
      Z_{\pi}\delta(\mathbf{r}_{\pi})
   \Bigr\rangle.
\end{equation}

\bibliography{pionic_helium}

\begin{table*}[t]
\caption{Nonrelativistic energies $E_{nr}$, Auger widths $\Gamma$, expectation values of operators ${\bf p}^4_e$, $\delta({\bf r}_{_{\rm He}})$, and $\delta({\bf r}_{\pi})$, and coefficients of the fine structure effective Hamiltonian $E_1$ for the $\pi{\rm ^4He^+}$ atom. The numerical precision on $E_{\rm nr}$ (indicated in parenthesis)
is better than $\sim 10^{-8}$, but the actual precision is limited to $>10^{-6}$ by the experimental
uncertainty on the $\pi^-$ mass used in these calculations (see text).\label{nonrela}}
\begin{center}
\begin{tabular}{c@{\quad}l@{\quad}l@{\quad}l@{\quad}l@{\quad}l@{\quad}l@{\quad}}
\hline\hline
\vrule width 0pt height 11pt 
state & $~~~~~~~~~~E_{nr}$ & $~~~~\Gamma/2$ & $~~~~~{\bf p}^4_e$
 & $~~\delta({\bf r}_{_{\rm He}})$ & $~~\delta({\bf r}_{\bar p})$ & $E_1$ \\
$(n,\ell)$ & ~~~~~~~~(a.u.) & ~~~(a.u.) &  & & & (MHz) \\
\hline\hline
\vrule width 0pt height 11pt 
$(15,14)$ & $-$3.0569481417(4)  &$5.1380\cdot10^{-6}$& 37.586951 & 1.271444 & 0.0807434 & $-$2244.04 \\
$(16,15)$ & $-$2.82854939373(4) & $2.1\cdot10^{-10}$ & 45.004106 & 1.495182 & 0.0606279 & $-$1954.99 \\
$(17,14)$ & $-$2.70984178(2)    & $2.00\cdot10^{-6}$ & 43.651419 & 1.423150 & 0.0308359 & $-$1143.79 \\
$(17,15)$ & $-$2.68542722(2)    & $2.50\cdot10^{-6}$ & 53.948888 & 1.762586 & 0.0402500 & $-$1521.19 \\
$(17,16)$ & $-$2.65751243850171 & $1.0\cdot10^{-13}$ & 52.830517 & 1.730041 & 0.0411226 & $-$1611.79 \\
$(18,15)$ & $-$2.58002554(1)    & $6.53\cdot10^{-6}$ & 60.776506 & 1.966939 & 0.0269818 & $-$1159.48 \\
$(18,16)$ & $-$2.556984919572(2)& $1.3\cdot10^{-11}$ & 60.537978 & 1.960761 & 0.0267217 & $-$1208.78 \\
$(18,17)$ & $-$2.5319465695913  & ---                & 60.487685 & 1.959449 & 0.0245532 & $-$1242.56 \\
$(19,15)$ & $-$2.50049802(7)    & $2.533\cdot10^{-5}$& 64.946166 & 2.090275 & 0.0184398 &  $-$867.61 \\
$(19,16)$ & $-$2.48154055239(1) & $2.0\cdot10^{-10}$ & 65.817187 & 2.119201 & 0.0184277 &  $-$913.16 \\
$(19,17)$ & $-$2.4618067856861  & ---                & 66.122061 & 2.128521 & 0.0164498 &  $-$919.72 \\
$(19,18)$ & $-$2.4413857971745  & ---                & 67.067533 & 2.156676 & 0.0127412 &  $-$895.64 \\
\hline\hline
\end{tabular}
\end{center}
\end{table*}

\begin{table*}
\caption{Nonrelativistic energies $E_{nr}$, Auger widths $\Gamma$, expectation values of operators ${\bf p}^4_e$, $\delta({\bf r}_{_{\rm He}})$, and $\delta({\bf r}_{\bar p})$, and coefficients of the fine structure effective Hamiltonian $E_1$ for the $\pi{\rm ^3He^+}$ atom. \label{nonrela2}}
\begin{center}
\begin{tabular}{c@{\quad}l@{\quad}l@{\quad}l@{\quad}l@{\quad}l@{\quad}l}
\hline\hline
state & $~~~~~~~~~~E_{nr}$ & $~~~~\Gamma/2$ & $~~~~~{\bf p}^4_e$
\vrule width 0pt height 11pt 
 & $~~\delta({\bf r}_{_{\rm He}})$ & $~~\delta({\bf r}_{\bar p})$ & $E_1$ \\
$(n,\ell)$ & ~~~~~~~~(a.u.) & ~~~(a.u.) &  & & & (MHz) \\
\hline\hline
\vrule width 0pt height 11pt 
$(15,14)$ & $-$3.0342533945(3)   &$5.5344\cdot10^{-6}$& 38.197137 & 1.289854 & 0.0789375 & $-$2247.04 \\
$(16,15)$ & $-$2.810277989054(3) & $3.47\cdot10^{-10}$& 45.712376 & 1.516455 & 0.0587005 & $-$1947.18 \\
$(17,14)$ & $-$2.695209116(1)    & $4.34\cdot10^{-7}$ & 50.220191 & 1.636126 & 0.0341126 & $-$1291.79 \\
$(17,15)$ & $-$2.6709980910(1)   & $6.0\cdot10^{-9}$  & 54.545365 & 1.780334 & 0.0386868 & $-$1504.51 \\
$(17,16)$ &$-$2.64312261030188(2)& $2.04\cdot10^{-12}$& 53.601086 & 1.753159 & 0.0392957 & $-$1593.92 \\
$(18,15)$ & $-$2.568490191(5)    & $5.845\cdot10^{-6}$& 61.363537 & 1.984751 & 0.0259702 & $-$1145.39 \\
$(18,16)$ & $-$2.545645472099(1) & $3.3\cdot10^{-11}$ & 61.166139 & 1.979672 & 0.0254795 & $-$1190.19 \\
$(18,17)$ & $-$2.52088142679     & ---                & 61.233513 & 1.981842 & 0.0230674 & $-$1218.10 \\
$(19,15)$ & $-$2.49096242(3)     & $1.552\cdot10^{-5}$& 65.831873 & 2.118160 & 0.0180330 &  $-$867.33 \\
$(19,16)$ & $-$2.47237023589(1)  & $2.2\cdot10^{-10}$ & 66.280943 & 2.133247 & 0.0176217 &  $-$897.91 \\
$(19,17)$ & $-$2.4529359745011   & ---                & 66.649182 & 2.144415 & 0.0155362 &  $-$900.79 \\
$(19,18)$ & $-$2.4329808449305   & ---                & 67.679405 & 2.175100 & 0.0117593 &  $-$870.97 \\
\hline\hline
\end{tabular}
\end{center}
\end{table*}

\begin{table*}
\caption{Transition energies and reduced matrix elements for transition amplitudes 
in $\pi{\rm ^4He}^+$ and $\pi{\rm ^3He}^+$ atoms. The energies include
relativistic and QED corrections of orders $R_{\infty}\alpha^2$ and $R_{\infty}\alpha^3$.
The fractional precision of the transition energies is limited to around $(3-5)\times 10^{-6}$, due to
the experimental uncertainty on the charged-pion mass used in the calculations (see the text).
}\label{transitions}
\begin{center}
\begin{tabular}{c@{\hspace{4mm}}c@{\hspace{6mm}}r@{\hspace{6mm}}r}
\hline\hline
 Atom & Transition & Frequency & Amplitude \\
  & $(n,\ell)\to(n^{\prime},\ell^{\prime})$    &     (GHz)       &   (a.u.) \\
\hline\hline
\vrule width 0pt height 11pt 
 $\pi{\rm ^4He^+}$ & $(16,15)\to(15,14)$ & 1502734.2 & 0.99894 \\
                 & $(16,15)\to(17,14)$ &  781052.6 & 0.11809 \\
                 & $(17,16)\to(16,15)$ & 1125306.1 & 1.47211 \\
                 & $(17,16)\to(18,15)$ &  509769.9 & 0.19859 \\
                 & $(18,16)\to(17,15)$ &  845055.5 & 1.79110 \\
                 & $(18,16)\to(19,15)$ &  371625.8 & 0.40616 \\
                 & $(18,17)\to(17,16)$ &  826119.9 & 2.05304 \\
                 & $(18,17)\to(19,16)$ &  331608.5 & 0.28201 \\
                 & $(19,17)\to(18,16)$ &  626195.2 & 2.32390 \\
                 & $(19,18)\to(18,17)$ &  595805.4 & 2.79140 \\
\hline
\vrule width 0pt height 11pt 
 $\pi{\rm ^3He^+}$ & $(16,15)\to(15,14)$ & 1473629.0 & 1.03276 \\
                 & $(16,15)\to(17,14)$ &  757070.5 & 0.12772 \\
                 & $(17,16)\to(16,15)$ & 1099765.9 & 1.51629 \\
                 & $(17,16)\to(18,15)$ &  490990.1 & 0.20566 \\
                 & $(18,16)\to(17,15)$ &  824726.1 & 1.83198 \\
                 & $(18,16)\to(19,15)$ &  359755.6 & 0.41722 \\
                 & $(18,17)\to(17,16)$ &  804244.8 & 2.11266 \\
                 & $(18,17)\to(19,16)$ &  319143.6 & 0.29336 \\
                 & $(19,17)\to(18,16)$ &  609953.1 & 2.37647 \\
                 & $(19,18)\to(18,17)$ &  578303.2 & 2.87201 \\
\hline\hline
\end{tabular}
\end{center}
\end{table*}

\end{document}